\pdfoutput=1
\documentclass[11pt,a4paper]{article}
\usepackage[margin=3cm]{geometry}
\usepackage{amsmath}
\usepackage{amssymb}
\usepackage{relsize}
\usepackage{xcolor}
\usepackage{hyperref}
\usepackage{url}
\usepackage{graphicx}
\usepackage{authblk} 
\usepackage{listings} 
\usepackage{cite} 
\usepackage{ulem}
\usepackage{slashed}
\usepackage{mathrsfs} 
\usepackage{subcaption}

\graphicspath{{./}{./figs/}}

\newcommand{\bd}{\texttt{BubbleDet}}
\newcommand{\ct}{\texttt{CosmoTransitions}}
\newcommand{\logphiinf}{$\log\phi_\infty$}
\newcommand{\tail}{\texttt{tail}}
\newcommand{\lpit}{\texttt{log\_phi\_inf\_tol}}
\newcommand{\docs}{\url{https://bubbledet.readthedocs.io/}}
\newcommand{\repo}{\url{https://bitbucket.org/og113/bubbledet/}}
\newcommand{\pypi}{\url{https://pypi.org/project/BubbleDet/}}
\newcommand{\conda}{\url{https://anaconda.org/conda-forge/bubbledet}}
\newcommand{\lb}{\overline{l}}
\newcommand\abs[1]{\left|#1\right|}
\newcommand{\te}{\textemdash}
\newcommand\MSbar{$\overline{\text{MS}}$}

\newcommand{\newtext}[1]{#1}
\newcommand{\mF}{m_{\text{F}}}

\title{\bd: A Python package to compute functional determinants for bubble nucleation}

\author{Andreas~Ekstedt\thanks{andreas.ekstedt@desy.de}}
\affil{Department of Physics and Astronomy, Uppsala University, P.O. Box 256, SE-751 05 Uppsala, Sweden}
\affil{II. Institute of Theoretical Physics, Universität Hamburg, D-22761, Hamburg, Germany}
\affil{Deutsches Elektronen-Synchrotron DESY, Notkestr. 85, 22607 Hamburg, Germany}

\author{Oliver Gould\thanks{oliver.gould@nottingham.ac.uk}}
\affil{School of Physics and Astronomy, University of Nottingham, Nottingham NG7 2RD, United Kingdom}

\author{Joonas Hirvonen\thanks{joonas.o.hirvonen@helsinki.fi}}
\affil{Helsinki Institute of Physics, University of Helsinki, FI-00014, Finland}

\date{29 August 2023}
\begin{document}
	\begin{flushright}
		\footnotesize
		{\large DESY-23-119} \\
	\end{flushright}
{\let\newpage\relax\maketitle}
	\thispagestyle{plain}
\abstract{We present a Python package \texttt{BubbleDet} for computing one-loop functional determinants around spherically symmetric background fields. This gives the next-to-leading order correction to both the vacuum decay rate, at zero temperature, and to the bubble nucleation rate in first-order phase transitions at finite temperature. For predictions of gravitational wave signals from cosmological phase transitions, this is expected to remove one of the leading sources of theoretical uncertainty. \texttt{BubbleDet} is applicable to arbitrary scalar potentials and in any dimension up to seven. It has methods for fluctuations of scalar fields, including Goldstone bosons, and for gauge fields, but is limited to cases where the determinant factorises into a product of separate determinants, one for each field degree of freedom. To our knowledge, \texttt{BubbleDet} is the first package dedicated to calculating functional determinants in spherically symmetric backgrounds}.
\clearpage

\section*{Program Summary}
\begin{description}
\item \textit{Program title:} \bd
\item \textit{Program obtainable from:} \pypi,\\ \conda
\item \textit{Documentation link:} \docs
\item \textit{Developer’s repository link:} \repo
\item \textit{Programming language:} Python 3
\item \textit{Distribution format:} Source (tar.gz) and built (.whl) distributions
\item \textit{Licence:} MIT
\item \textit{Operating system:} Compatible with any OS with Python installed. Tested on Linux (Ubuntu 20 and 22), macOS Ventura 13.1 and Windows 10.
\item \textit{External routines:} CosmoTransitions \cite{Wainwright:2011kj}, FinDiff \cite{findiff}, NumPy \cite{harris2020array}, SciPy \cite{2020SciPy-NMeth}
\item \textit{RAM:} 8-10 MB for our examples.
\item \textit{Typical running time:} 0.25 seconds on a laptop with a 3.2 GHz processor for a typical bubble, growing to a few seconds for a bubble in the thin wall regime.
\item \textit{Nature of problem:} The problem is to efficiently compute functional determinants for tunnelling rates in quantum field theory. It is assumed that the background field has spherically symmetry, as is relevant to vacuum decay or bubble nucleation.
\item \textit{Solution method:} The functional determinant is decomposed into spherical harmonics. For each value of the total angular momentum, we use the Gelfand-Yaglom theorem to compute the reduced functional determinant in terms of an initial value problem. Zero eigenvalues are treated using a result of Dunne and Min, which is equivalent to the collective coordinate method. The sum over angular momenta is regularised in the \MSbar\ scheme, and convergence of the sum is accelerated using the WKB approximation.
\item \textit{Restrictions:} The code is currently limited to bosonic fields, and to cases where the functional determinant factorises into a product of separate determinants, one for each field degree of freedom.
\end{description}
\clearpage


\section{Introduction} \label{sec:Introduction}

Functional determinants are ubiquitous in field theory, encoding the effects of quantum or statistical fluctuations. Starting from the path integral, they arise whenever one makes a semiclassical or saddlepoint approximation, and hence appear in a wide range of physical phenomena.
For example, functional determinants play a central role in vacuum decay \cite{Callan:1977pt, Isidori:2001bm, Andreassen:2017rzq}, thermal bubble nucleation \cite{Langer:1967ax, Affleck:1980ac, Linde:1981zj}, the study of solitons \cite{Dashen:1974ci}, and baryon number violation through anomalies \cite{tHooft:1976snw, Arnold:1987mh}.

The semiclassical or saddlepoint approximation to a path integral is the infinite-dimensional generalisation of Laplace's method. For a set of fields $\phi_i(x)$ with action $S[\phi_i]$, this takes the form
\begin{equation} \label{eq:intro_saddlepoint}
\int \mathcal{D} \phi_i\ e^{-S[\phi_i]} \approx A\ e^{-B}.
\end{equation}
The largest contribution to the result, $B$, is equal to the action evaluated at a saddlepoint. The prefactor $A$ comes from fluctuations around the saddle, which at leading order arise quadratically in the action. Effectively the path integral becomes an infinite product of Gaussian integrals that, in turn, results in an infinite product of eigenvalues\te a functional determinant \cite{Langer:1967ax, Callan:1977pt}.

The computation of the saddlepoint action $B$ has received much attention. For vacuum decay, many different algorithms have been proposed \cite{Coleman:1977py, Espinosa:2018hue, Jinno:2018dek, Piscopo:2019txs, Chigusa:2019wxb, Bardsley:2021lmq}, and there are at least five software packages dedicated to this computation \cite{Wainwright:2011kj, Masoumi:2016wot, Athron:2019nbd, Sato:2019wpo, Guada:2020xnz}. On the other hand, to our knowledge there are no existing public codes capable of computing the functional determinant.

The computation of $A$ is the focus of this article. Typically, it has been assumed that $A$ is of negligible numerical importance in comparison with $e^{-B}$, and hence one may be content with an estimate for $A$ based on dimensional analysis. This assumption is often motivated by the exponential form of equation~\eqref{eq:intro_saddlepoint}, and the familiar rapidity of exponential growth.

Despite appearances, the functional determinant $A$ generically takes an exponential form in field theory. While there is undoubtedly some arbitrariness in the distinction between exponential and prefactor, below we argue that field-theoretic functional determinants naturally give exponentials.
\begin{itemize}
\item The prefactor is the sum of one-loop vacuum Feynman diagrams, including disconnected diagrams. This is equal to the exponential of the sum of connected vacuum diagrams, $A=e^{\text{connected}}$. The divergences of the connected diagrams cancel against the one-loop counterterms in the action, leaving a finite remainder.
\item The prefactor is the integral over the phase space of fluctuations around a saddlepoint, and hence it is the exponential of their entropy. For a field theory, these fluctuations are an infinite number of harmonic oscillators. The entropy of a single oscillator is $S_i=\mathcal{O}(1)$, and the total entropy is extensive: $A = e^{\sum_i S_i}$.
\item Consider a simple example, such as a constant background field. In this case $B=\int_x V_{0}$ and $A=e^{-\int_x V_{1}}$, where $V_{0}$ and $V_{1}$ are the tree-level potential and one-loop effective potential respectively.
\end{itemize}
As a consequence of this exponentiation, after scaling out dimensions, the typical magnitude of $\log A$ relative to $B$ is the same as for any other one-loop correction to the tree-level.%
\footnote{We use $\log$ to refer to the natural logarithm throughout, following NumPy, SciPy etc.}
Furthermore, for phase transitions the tree-level action is typically fine-tuned, thus making the relative impact of $A$ even greater. It is however important to stress that perturbation theory may still work since higher loop corrections are suppressed by additional powers of couplings, and it is but the tuned tree-level action at fault.

To date there have been several calculations of functional determinants in various field-theoretic models. Analytic results have been attained in one spatial dimension \cite{Dashen:1974cj}, in the thin-wall limit \cite{Konoplich:1987yd, Munster:1999hr, Munster:2000qt, Munster:2003an, Garbrecht:2015oea, Ivanov:2022osf}, and in a scaleless potential \cite{Isidori:2001bm, Andreassen:2017rzq}. Outside these simplifying cases, computations have been carried out numerically \cite{Baacke:1993ne, Baacke:1993aj, Baacke:1995bw, Baacke:2003uw, Dunne:2004cp, Dunne:2006ct, Baacke:2008zx, Ekstedt:2021kyx}. Calculations of functional determinants are universally involved, and consequently they have yet to achieve widespread usage, either for phenomenological applications or for comparisons to new theoretical methods. Motivated by this, we introduce \bd, a Python package that automatically calculates bosonic functional determinants in spherically symmetric scalar field backgrounds.

\section{Quick start} \label{sec:QuickStart}

The simplest way to install \bd\ is to use a Python package manager, such as the Package Installer for Python (PIP) or Conda. To install with PIP, run the following in a Linux or Unix (including macOS) terminal or in a Windows command prompt%
\footnote{Use \texttt{pip3} in place of \texttt{pip} on systems where \texttt{pip} refers to the \texttt{Python 2} instance.}
\begin{lstlisting}[language=bash]
$ pip install BubbleDet
\end{lstlisting}
Alternatively, to install with Conda, use
\begin{lstlisting}[language=bash]
$ conda install -c conda-forge BubbleDet
\end{lstlisting}
Once installed, \bd\ can be imported just as any other Python package.

To use \bd\ to compute functional determinants, one must pass a precomputed bounce solution. Since \bd\ is written in Python, it is straightforward to obtain and pass the bounce solution from \ct\ \cite{Wainwright:2011kj}. This is however optional, and \bd\ can be used together with any bounce solver.

A number of examples are included with the package, the simplest of which, \texttt{first\_example.py}, demonstrates the computation of the functional determinant for the vacuum decay of a pure scalar field with a quartic potential. Further examples demonstrate how to use the package in a variety of settings, including for a symmetry-breaking vacuum transition in the Abelian Higgs model and for a high-temperature phase transition in a Yukawa model. All the examples can be viewed in the online documentation, at \docs, where one can also find comprehensive documentation of all the package functions.

\section{Theoretical background} \label{sec:Num}

\subsection{Overview of the problem}\label{sec:NumOverview}

A metastable or false vacuum state can decay through bubble nucleation. This is a semiclassical process, dominated by a critical bubble or bounce, which leads to a local escape from the false vacuum. In quantum field theory, the study of this process was initiated by Coleman and Callan in the late 70s \cite{Coleman:1977py, Callan:1977pt}, building on earlier work in classical statistical field theory \cite{Langer:1967ax, Langer:1969bc}. 

More recently, the predicted metastability of the electroweak vacuum state has necessitated the development of quantitatively reliable predictions for the decay rate \cite{Isidori:2001bm, Andreassen:2017rzq}. This has been bolstered by the possibility of observing gravitational waves from a cosmological phase transition in the early universe \cite{Witten:1984rs}, for which accurate predictions of the bubble nucleation rate are required to determine the peak frequency and amplitude of the signal \cite{Croon:2020cgk, Gould:2021oba}.

\begin{figure}[t]
    \centering
    \includegraphics[width=0.4\textwidth]{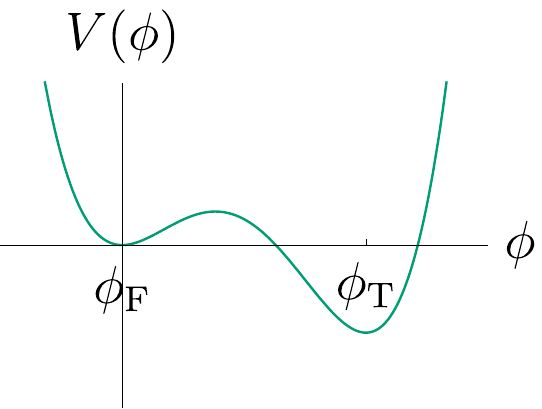}
    \caption{Schematic example potential showing a (metastable) false vacuum at $\phi_\text{F}$ and a (stable) true vacuum at $\phi_\text{T}$.}
    \label{fig:PotentialIntro}
\end{figure}
Consider a single real scalar field in $d$-dimensions with Euclidean action
\begin{align}
S[\phi]=\int d^dx\left[\frac{1}{2}\nabla_\mu \phi \nabla_\mu \phi+V(\phi)\right],
\end{align}
where we have kept all indices lowered to emphasise the Euclidean signature of the metric. Let us assume that $V(\phi)$ has one (metastable) minimum at $\phi=\phi_\text{F}$ and a deeper one at $\phi=\phi_\text{T}$, such as in Figure \ref{fig:PotentialIntro}. The metastable, or false, vacuum will then decay to the stable, or true, vacuum with a rate per unit volume given at zero temperature by~\cite{Coleman:1977py, Callan:1977pt, Vainshtein:1981wh, Andreassen:2016cvx, Andreassen:2017rzq, Ai:2019fri}
\begin{align}\label{eq:DecayRate}
\Gamma =
\left(\frac{S[\phi_\text{b}]}{2\pi}\right)^{d/2}
\left\vert
\frac{\det'\left(-\nabla^2+V''(\phi_\text{b})\right)}{\det\left(-\nabla^2+V''(\phi_\text{F})\right)}
\right\vert^{-1/2}
e^{-(S[\phi_\text{b}]-S[\phi_\text{F}])}.
\end{align}
Here $\det'$ signifies that the translational zero modes are omitted from the determinant, and $\phi_\text{b}$ denotes the bounce solution, also referred to as the critical bubble.

We assume that the bounce only depends on the radial coordinate $r\equiv \sqrt{x_\mu x_\mu}$ and satisfies~\cite{Coleman:1977py}
\begin{align}\label{eq:BounceEq}
&\frac{\delta S[\phi]}{\delta \phi}\bigg|_{\phi_\text{b}} = -\partial^ 2\phi_\text{b}(r)-\frac{(d-1)}{r} \partial \phi_\text{b}(r) + V'(\phi_\text{b}) = 0,
\end{align}
subject to the boundary conditions,
\begin{align}
    \lim_{r \rightarrow \infty} \phi_\text{b}(r)=\phi_\text{F},
    \qquad \left. \partial \phi_\text{b}\right\vert_{r=0}=0.
\end{align}
Here we have introduced the shorthand $\partial\equiv \frac{\partial}{\partial r}$.
These boundary conditions are preserved by the functions included in the functional determinant, as they can be considered additive fluctuations about the background. The fluctuations are therefore regular at the origin and go to zero at infinity.

Note that equation \eqref{eq:DecayRate} only incorporates 1-loop corrections\te via the determinant\te and higher loop corrections are omitted in this work.

\paragraph{Multi-field models}
In more complicated models, the full functional determinant runs over all the fields. In principle these fields can mix, either through the mass matrix, or through derivative terms. However, in the present work we will assume that the functional determinants can be diagonalised in field space. We also assume that that there is only one background field $\phi(r)$ that is coordinate dependent.

As a concrete example, consider first a scalar theory with a global U(1) symmetry and with Euclidean Lagrangian
\begin{align} \label{eq:U1GlobalLagrangian}
   \mathscr{L}&=
   (\nabla_\mu \Phi)^*  \nabla_\mu\Phi + V(\Phi), \\
V(\Phi) &= m^2 \Phi^*\Phi + \lambda (\Phi^*\Phi)^2.
\end{align}
If we now expand around a radially-symmetric background $\phi(r)$ that extremises the action, $\Phi=\frac{1}{\sqrt{2}}\left(\phi(r) + H(x)+ i G(x)\right)$, we find
\begin{align} \label{eq:U1GlobalExpanded}
\mathscr{L}=
&\mathscr{L}(\phi)
+\frac{1}{2}H \underbrace{\left[-\nabla^2+V''(\phi)\right]}_{\equiv \mathcal{O}_{H}(\phi)}H 
+\frac{1}{2}G\underbrace{\left[-\nabla^2+\phi^{-1}V'(\phi)\right]}_{\equiv \mathcal{O}_{G}(\phi)} G
+ \dots
\end{align}
where the dots denote terms higher order than quadratic in fluctuations. In this case, the quadratic part of the Lagrangian, and consequently the functional determinant, factorises into Higgs and Goldstone pieces.

The generalization of equation \eqref{eq:DecayRate} to this model is then
\begin{align} \label{eq:U1GlobalRate}
\Gamma =
\mathcal{J}_{G} \mathcal{V}_{G}
\sqrt{\frac{\det\mathcal{O}_{G}(\phi_\text{F})}{\det'\mathcal{O}_{G}(\phi_\text{b})}}
\mathcal{J}_{H}
\sqrt{\left|\frac{\det\mathcal{O}_{H}(\phi_\text{F})}{\det'\mathcal{O}_{H}(\phi_\text{b})}\right|}
    e^{-(S[\phi_\text{b}]-S[\phi_\text{F}])},
\end{align}
where $\mathcal{V}_X$ and $\mathcal{J}_{X}$ are volume and Jacobian factors arising due to zero modes. The former is the volume of the space of zero modes, and the latter is the Jacobian for the transformation to collective coordinates \cite{GERVALS1976281,Vainshtein:1981wh}. They are discussed further in Section \ref{sec:NumOneLoop} and in Appendix \ref{app:Volume}. For the Higgs, we have that $\mathcal{J}_{H} = \left(S[\phi_\text{b}]/(2\pi)\right)^{d/2}$. Note that we need not take the modulus of the Goldstone determinant, as only the Higgs determinant contains a negative mode.

Extending this model, consider next the inclusion of an $n$-component scalar field, coupled as
\begin{align}
    \Delta \mathscr{L} = \frac{1}{2}\nabla_\mu \chi^a \nabla_\mu \chi^a + \frac{1}{2}M^2 \chi^a\chi^a + \kappa \chi^a\chi^a \Phi^*\Phi ,
\end{align}
where the index $a$ runs over $1, 2, \dots, n$.
Assuming the $\chi$ field does not take a background expectation value, the Lagrangian expanded to quadratic order reads
\begin{align}
    \Delta \mathscr{L} = \frac{1}{2}\chi^a
    \underbrace{\left[ - \nabla^2 + M^2 + \kappa \phi^2 \right]}_{\equiv \mathcal{O}_\chi(\phi)}
    \chi^a + \dots
\end{align}
There are no zero modes for the $\chi$ field, as it does not break any symmetry. It also does not mix with the Higgs or Goldstone fluctuations at quadratic order. So, the full nucleation rate reads
\begin{align} \label{eq:TwoScalarRate}
\Gamma =
\left(\frac{\det\mathcal{O}_{\chi}(\phi_\text{F})}{\det\mathcal{O}_{\chi}(\phi_\text{b})}\right)^{n/2}
\mathcal{J}_{G} \mathcal{V}_{G}
\sqrt{\frac{\det\mathcal{O}_{G}(\phi_\text{F})}{\det'\mathcal{O}_{G}(\phi_\text{b})}}
\mathcal{J}_{H}
\sqrt{\left|\frac{\det\mathcal{O}_{H}(\phi_\text{F})}{\det'\mathcal{O}_{H}(\phi_\text{b})}\right|}
    e^{-(S[\phi_\text{b}]-S[\phi_\text{F}])}.
\end{align}

In all the examples above, the total one-loop contribution to the decay rate is a product of independent functional determinants, each of which takes the form
\begin{align} \label{eq:WIntro}
    \mathcal{J} \mathcal{V} \left|\frac{\det'(-\nabla^2 + W(r))}{\det(-\nabla^2 + W(\infty))}\right|^{-n/2},
\end{align}
where $n$ is the number of such field degrees of freedom, and $\mathcal{J}$ and $\mathcal{V}$ are the Jacobian and volume factors for the zero modes. Since we are interested in the rate per unit volume, we always remove the volume associated with space-time translations.

The $W$ factor in equation \eqref{eq:WIntro} denotes a field-dependent mass squared, for example
\begin{align}
    W(r) =
    \begin{cases}
        V''(\phi_\text{b}(r)), & \text{Higgs}, \\
        \phi_\text{b}(r)^{-1} V'(\phi_\text{b}(r)), & \text{Goldstone}, \\
        M^2 + \kappa \phi_\text{b}(r)^2, & \chi\text{ field.}
    \end{cases}
\end{align}
For the above models, a list of $W$ functions determines the Lagrangian at quadratic order. However, in models with off-diagonal terms at quadratic order, $W$ should be promoted to a matrix in field space.

The inclusion of vector fields, such as in the electroweak theory, inevitably leads to off-diagonal terms in field space, through mixing with Goldstone fields \cite{Andreassen:2017rzq, Endo:2017tsz, Ai:2020sru, Ekstedt:2021kyx}. Accounting for such mixing terms goes beyond the scope of this first version of \bd. However, one can approximate the full vector-field one-loop determinant by dropping the off-diagonal terms. This can be expected to capture the correct order of magnitude of the functional determinant. Vector fields are discussed further in Appendix \ref{app:Vector}. \newtext{Nevertheless, while we do not consider full multi-field determinants in this work, several such computations exist in the literature~\cite{Isidori:2001bm, Andreassen:2017rzq, Chigusa:2018uuj, Chigusa:2022xpq, Chigusa:2020jbn,Ai:2018guc,Ekstedt:2021kyx}. }

\paragraph{Finite temperature transitions}
At high temperatures it is possible for fields to borrow energy from the thermal bath and escape from a metastable state.  In addition, if the nucleating field evolves parametrically slower than the inverse temperature, we can describe its dynamics classically in real time. As such, J.S.~Langer's framework of classical nucleation theory is applicable \cite{Langer:1969bc, Langer:1967ax, Gould:2021ccf, Ekstedt:2022tqk}, and the rate factorises into dynamical and statistical parts
\begin{align} \label{eq:ThermalRate}
\Gamma_T = A_\text{dyn} \times A_\text{stat}.
\end{align}

For thermal nucleation in $d+1$ spacetime dimensions, the statistical factor $A_\text{stat}$ coincides with the vacuum decay rate in $d$ dimensions given by equation \eqref{eq:DecayRate}, and thus can be computed directly with \bd. Albeit with one caveat: Thermal corrections from nonzero Matsubara modes should be included in the tree-level potential when computing $A_\text{stat}$~\cite{langer1974metastable, Gould:2021ccf, Linde:1981zj}.

The dynamical factor $A_\text{dyn}$ contains dissipative effects and, unlike $A_\text{stat}$, it is not expected to exponentiate.
In Langer's framework, which assumes Langevin dynamics, the dynamical factor is equal to the real-time growth rate of a critical bubble divided by $2\pi$ . This can be expressed as \cite{Langer:1969bc, Hanggi:1990zz, Berera:2019uyp}
\begin{align} \label{eq:DynamicalPrefactor}
A_\text{dyn} = \frac{1}{2\pi}\left(\sqrt{|\lambda_-| + \frac{\eta^2}{4}} - \frac{\eta}{2}\right),
\end{align}
in terms of the negative eigenvalue of the functional determinant $\lambda_-$, and the Langevin damping coefficient $\eta$. The negative eigenmode is the lowest eigenvalue of the Higgs fluctuation operator $\mathcal{O}_H(\phi_\text{b})$, and corresponds to isotropic growth or shrinking of the bubble.
This identification receives corrections at higher orders \cite{Ekstedt:2022tqk}.
The computation of $\lambda_-$ can be carried out using \bd, but that of $\eta$ requires additional real-time input. Setting $\eta = 0$ yields the approximation of Ref.~\cite{Affleck:1980ac}, though in this limit the saddlepoint approximation is expected to break down \cite{Hanggi:1990zz}.

\subsection{The Gelfand-Yaglom theorem}\label{sec:NumGelYagThm}
To find the rate in equation \eqref{eq:DecayRate} we need to evaluate the functional determinant $\det\left(-\nabla^2+W(r)\right)$. For a constant scalar field $\phi$ one finds the usual effective potential~\cite{Coleman:1973jx,PhysRevD.9.1686}, however, closed analytical expressions are in general not available for a spatially varying field.

Instead numerical techniques are required. As an initial step it is useful to exploit the spherical symmetry and to expand all eigenfunctions in spherical harmonics:
\begin{align}\label{eq:DetPartialWaves}
	\frac{\det\left(-\nabla^2+W(r)\right)}{\det\left(-\nabla^2+W(\infty)\right)}=\mathlarger{\Pi}_{l=0}^{\infty} \left[\frac{\det\left(-\nabla_l^2+W(r)\right)}{\det\left(-\nabla_l^2+W(\infty)\right)}\right]^{\deg(d;l)},
\end{align}
where
\begin{align}
 \deg(d;l)=\frac{(d+2 l-2) \Gamma (d+l-2)}{\Gamma (d-1) \Gamma (l+1)}
\end{align}
 is the degeneracy factor for the orbital quantum number $l$. Dependence on the magnetic orbital quantum number (normally denoted $m$) is trivial; it is completely accounted for by the degeneracy factor. The spherical Laplacian is
 \begin{align}
 	\nabla_l^2=\partial^2+\frac{d-1}{r}\partial -\frac{l(l+d-2)}{r^2}.
 \end{align}

To compute the determinant for given $l$ we use the Gelfand-Yaglom theorem, which in our case states that~\cite{Gelfand:1959nq,Forman1987,Kirsten:2010eg,Kirsten:2004qv}
\begin{align}
\frac{\det\left(-\nabla_l^2+W(r)\right)}{\det\left(-\nabla_l^2+W(\infty)\right)}=\frac{\psi^l_\text{b}(\infty)}{\psi^l_\text{F}(\infty)},
\end{align}
where the $\psi^l_{b,F}(r)$ satisfy the differential equations
\begin{align}\label{eq:GYDiffEq}
\left[	-\nabla_l^2+W(r)\right]\psi^l_{b}(r)&=0,
& \left[	-\nabla_l^2+W(\infty)\right]\psi^l_{F}(r)&=0,
\end{align}
with the boundary condition $\psi^l_{b,F}(r)\sim r^{l}$ as $r \rightarrow 0$. Note that these equations for $\psi^l_{b,F}$ are initial value problems, whereas the corresponding eigenfunctions satisfy boundary value problems. Since $W(\infty)$ is a constant, the differential equation for $\psi^l_{F}(r)$ can be solved analytically.

\subsection{One-loop correction to the action}\label{sec:NumOneLoop}
As discussed in Section \ref{sec:NumGelYagThm}, the problem of calculating the rate in equation \eqref{eq:DecayRate} is reduced to solving the differential equations
\begin{align}\label{eq:repeatedpsi}
 \left[-\nabla_l^2+W(r)\right] \psi^l_{b}(r)=0,
\end{align}
for each $l$. Given the bounce, $\phi_\text{b}(r)$, these equations can be solved numerically. There are however a few complications.

First, the determinant vanishes if there are zero modes, so these have to be removed. Second, in practice we can only solve equation \eqref{eq:GYDiffEq} for a finite number of $l$'s. And third, the product\te or equivalently a sum in the exponent\te in equation \eqref{eq:DetPartialWaves} is generically ultraviolet divergent. Let us deal with these problems in turn. 

\paragraph{Zero modes}
If we have a pure scalar theory, all zero modes occur either in the $l=0$ or in the $l=1$ determinant. The procedure to remove zero modes is equivalent for the two cases so we focus on the latter, and refer to Appendix \ref{app:GoldstoneZeroModes} for the $l=0$ case. For a single scalar, equation \eqref{eq:repeatedpsi} with $l=1$ gives
\begin{align} \label{eq:l1DifferentialEquation}
\left[-\partial^2-\frac{d-1}{r}\partial +\frac{(d-1)}{r^2}+V''(\phi_\text{b})\right]\psi^1_{b}(r)=0,
\end{align}
which has the solution $\psi^1_{b}(r)\propto \partial \phi_\text{b}(r)$. Note that the determinant indeed vanishes since $\lim_{r\rightarrow \infty} \partial \phi_\text{b}(r)=0$.

Formally one can remove these zero modes\te which arise because the bounce breaks the translation symmetry\te by using collective coordinates~\cite{GERVALS1976281,Vainshtein:1981wh}. This means that we re-express eigenfunctions that generate the symmetry as a coordinate shift for all other eigenfunctions. Then, since everything is transitionally invariant, the integration over all possible translations gives the $d$-dimensional volume $\mathcal{V}$; we also have to include a Jacobian factor, $\mathcal{J}$, since we changed variables. Our job now is to find the value of the determinant once zero modes have been removed.

To do this we follow~\cite{Baacke:1993ne, Baacke:2003uw, Dunne:2005rt, Dunne:2006ct} and deform the equation to
\begin{align} \label{eq:l1DeformedDifferentialEquation}
	\left[-\partial^2-\frac{d-1}{r}\partial +\frac{(d-1)}{r^2}+V''(\phi_\text{b})+k^2\right]\psi^{1,k}_{b}(r)=0,
\end{align}
which effectively shifts all eigenvalues by $k^2$. The determinant without zero modes is then reproduced by the following limit
\begin{align}
\frac{\det'\left(-\nabla_1^2+V''(\phi_\text{b})\right)}{\det\left(-\nabla_1^2+V''(\phi_\text{F})\right)}=\lim_{k\rightarrow 0}  \frac{\psi^{1,k}_\text{b}(\infty)}{k^2}\frac{1}{\psi^1_\text{F}(\infty)}.
\end{align}
After taking the $k\rightarrow 0$ limit one finds~\cite{Dunne:2006ct}
\begin{align}\label{eq:packTransZeroModes}
\underbrace{\left(\frac{S[\phi_\text{b}]}{2\pi}\right)^{d/2}}_{=\mathcal{J}}	\left(\frac{\det'\left(-\nabla_1^2+V''(\phi_\text{b})\right)}{\det\left(-\nabla_1^2+V''(\phi_\text{F})\right)}\right)^{-d/2}
=
\left[(2\pi)^{d/2-1}\phi_\infty \abs{\partial^2\phi_\text{b}(0)}\right]^{d/2}.
\end{align}
Here $\phi_\infty$ is defined from the asymptotic behaviour of $\phi_\text{b}(r)$ as $r\rightarrow \infty$ via
\begin{align} \label{eq:phi_infinity}
\phi_\text{b}(r)\sim \phi_{\text{F}} + \phi_\infty\,\text{K}(d/2-1,\mF r)\, \left(\frac{\mF}{r}\right)^{d/2-1}.
\end{align}
In addition, from equation \eqref{eq:BounceEq} we find that $\partial^2\phi_\text{b}(0)= \frac{1}{d}\frac{dV}{d\phi}(\phi_\text{b}(0))$.

For potentials in which the nucleating, or Higgs, field is massless in the metastable phase, $W(\infty)=0$, the calculation in~\cite{Dunne:2005rt,Dunne:2006ct} needs to be modified. In this case the asymptotic behaviour of the bounce follows from the $\mF\to 0_+$ limit of equation \eqref{eq:phi_infinity},
\begin{align}
    \phi_\text{b}(r)\sim \phi_{\text{F}} + \phi_\infty\, \frac{\Gamma\left(d/2-1\right)2^{d/2-2} }{r^{d-2}},
\end{align}
here assuming $d>2$.
\footnote{In $d=2$ the behavior of the massless asymptotic solution is more complicated. We do not consider this case.}
While the derivation differs from the massive case, the final result for the determinant agrees with equation \eqref{eq:packTransZeroModes}. It is worked out in Appendix \ref{app:massless_zero_modes}.

\paragraph{Analytic solution for large $l$}
For asymptotically large $l$, the computation of the determinant simplifies.
Physically we can think of $l$ as the classical orbital momentum $l\sim \vec{p}\times \vec{r}$; this means that the source at the origin\te the critical bubble background\te becomes less significant when $l\gtrsim m R$, where $R$ is the bubble radius and $m$ is the particle mass. So for large $l$ we can use a WKB approximation to solve equation \eqref{eq:GYDiffEq} analytically~\cite{Dunne:2004cp}. To do this we use R.E.~Langer's method~\cite{Langer:1937qr} and define $\Psi(x)=r^{d/2-1}\psi(r)$ together with a change of variables to $x=\log r$. Equation \eqref{eq:GYDiffEq} is then equivalent to
\begin{align}
	\partial_x^2\Psi^l_{b,F}(x) = A^2_{b,F}(x)\Psi^l_{b,F},
    \quad A^2_{b,F}(x) = e^{2x}W(e^x)+\lb^2,\quad \lb \equiv l + \frac{d-2}{2}.
\end{align}
For large $l$ we can solve this equation in powers of $l^{-1}$. We leave the details to Appendix \ref{app:WKB} and merely quote the leading order solution~\cite{Dunne:2004cp}
\begin{align}\label{eq:GYLOWKB}
& \log \frac{\psi^l_{b}(\infty)}{\psi^l_{F}(\infty)}
= \frac{1}{2 \lb} \int \mathrm{d}r r\Delta W(r)+\mathcal{O}\left(l^{-3}\right),
\end{align}
where we have defined
\begin{align}\label{eq:DeltaW}
    \Delta W (r) & \equiv W(r) - W(\infty).
\end{align}

In practice we use the Gelfand-Yaglom theorem to solve equation \eqref{eq:GYDiffEq} numerically for $l$ up to some $l_\text{max}$. We then use the WKB approximation to solve equation \eqref{eq:GYDiffEq} analytically for all remaining $l$'s.

\paragraph{Divergent sums and renormalization}
From equation \eqref{eq:GYLOWKB} we see that for large $l$
\begin{align}
\log \left[\frac{\det'\left(-\nabla_l^2+W(r)\right)}{\det\left(-\nabla_l^2+W(r)\right)}\right]^{\deg(d;l)}
\approx 
\deg(d;l) \left[\frac{1}{2 \lb} \int \mathrm{d}r r \Delta W (r)+\mathcal{O}\left(l^{-3}\right)\right] \nonumber
\end{align}
The degeneracy factor scales as $l^{d-2}$, so in general there is a divergent sum for all $d\geq2$.

Take for example $d=4$. In this case we also need $\mathcal{O}\left(l^{-3}\right)$ terms:
\begin{align}\label{eq:WKBDiv}
\log \frac{\psi^l_{b}(\infty)}{\psi^l_{F}(\infty)}
\approx
&\frac{1}{2 \lb} \int \mathrm{d}r r \Delta W (r)
-\frac{1}{8 \overline{l}^3}\int \mathrm{d}r r^3\left[W(r)^2-W(\infty)^2\right].\nonumber
\end{align}
Because the degeneracy factor is $\deg(4;l)=(l + 1)^2$, both terms diverge once we sum over $l$. To handle this we use dimensional regularization and set $d=4-2\epsilon$. The only sum with a pole is of the form
\begin{align}
	\sum_{l=2}^{\infty}\deg(d;l) \overline{l}^{-3}=\frac{1}{2\epsilon}+\mathcal{O}(\epsilon).
\end{align}
In the $\overline{\text{MS}}$-scheme we should multiply this expression by $\left(\frac{\exp (\gamma ) \mu ^2}{4 \pi }\right)^{\epsilon }$ and add counter-terms\te these can directly be read off from the (constant-field)%
\footnote{For $d>4$ we must also include counter-terms that depend on spatial derivatives of $W$. These can be found from sub-leading terms in the derivative expansion, see for example~\cite{PhysRevD.46.1671}.}
Coleman-Weinberg potential~\cite{Coleman:1973jx} :
\begin{align}
V_\text{eff}(\phi) = V(\phi)+V_\text{ct}(\phi)+V_1(\phi),
\quad V_1(\phi) = -\frac{C}{4(4\pi)^2\epsilon}W(\phi)^2+\ldots,
\end{align} 
which fixes $V_\text{ct}(\phi)$. Here $C=1$ for Higgs and Goldstone determinants and $C=d-1$ for vector determinants.  After integrating over the volume we find
\begin{align} \label{eq:CountertermAction}
S_\text{ct}[\phi] = \frac{\left.C\right\vert_{\epsilon=0}}{32 \epsilon}\int \mathrm{d}r  r^{3-2 \epsilon } \frac{\pi ^{-\epsilon }}{\Gamma (2-\epsilon )} W(\phi)^2.
\end{align}
Note that this contribution enters the rate as $e^{-S_\text{ct}[\phi]}$.

Putting everything together and using
\begin{align}
\mathlarger{\mathlarger{\Pi_l}} \left[ \frac{\psi^l_{b}(\infty)}{\psi^l_{F}(\infty)} \right]^{-\deg(d;l)/2}=\exp\left[-\frac{1}{2} \sum_l \deg(d;l)\log \frac{\psi^l_{b}(\infty)}{\psi^l_{F}(\infty)}\right],
\end{align}
gives
\begin{align}\label{eq:RenormalizedSum}
-\frac{C}{16}\int \mathrm{d}r rr^3\left[W(\phi_{b}(r))^2-W(\phi_{F})^2\right]\left[\log \left(\frac{\mu r}{2}\right)-a-\frac{1}{2}+\gamma\right],
\end{align}
where $(C,a)=(1,0)$ for scalars, and $(C,a)=(d-1, 1/(d-1))$ for vectors.

Similar terms appear in all even-numbered dimensions, albeit they multiply different terms in the WKB approximation. To renormalize our theory in $d$ dimensions we need all terms up to $l^{-d+1}$ in the WKB approximation.%
\footnote{In some dimensions, for example in $d=6$, $\deg(6;l)\sim \frac{1}{12}l^4-\frac{1}{12}l^2$, there can be several  poles. That is, when the $l^4$ and $l^2$ terms combine with the $l^{-5}$ and $l^{-3}$ terms in the WKB expansion.}

\section{The \bd\ code} \label{sec:Code}

\begin{figure}[t]
    \centering
    \includegraphics[width=1.0\textwidth]{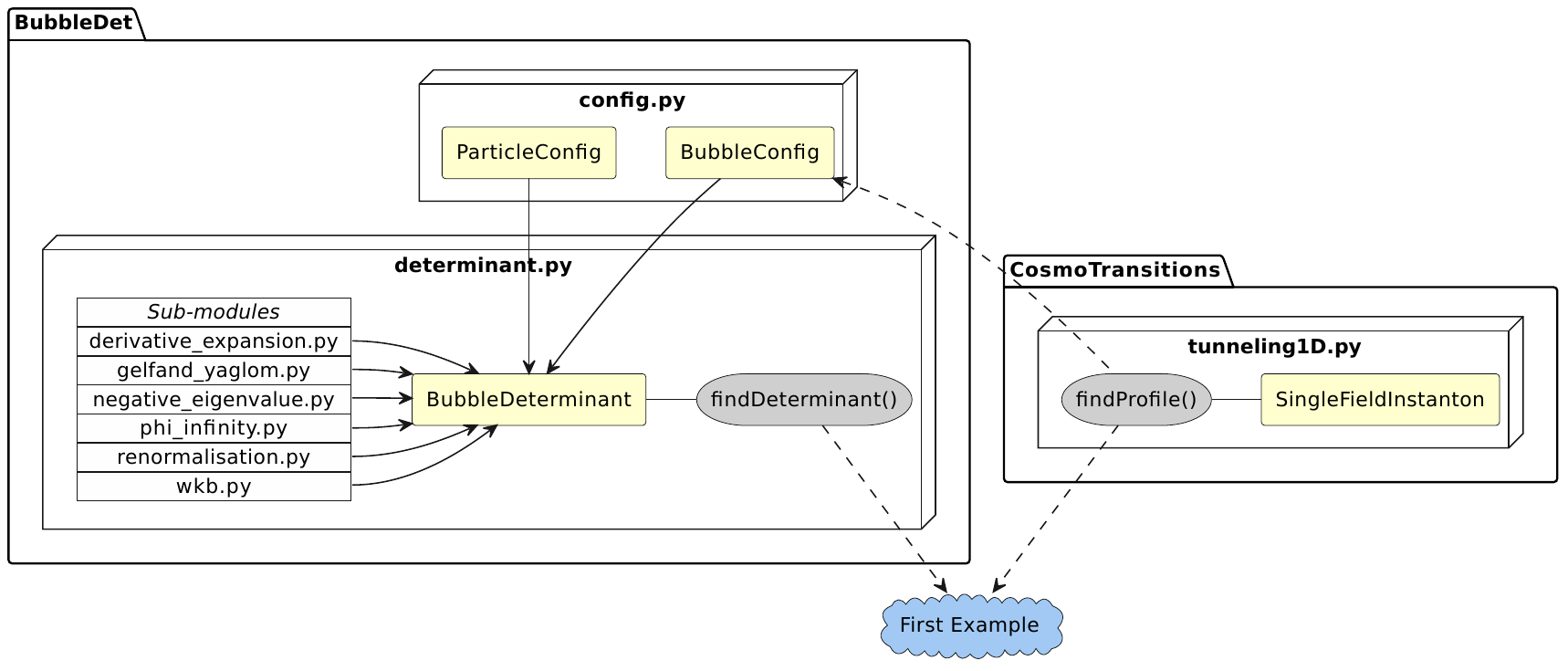}
    \caption{Component diagram of the \bd\ package, here shown when \bd\ is used together with \ct. The full arrows show a strong connection between components which is structurally essential, while the dashed arrows show a weaker connection, such as a particular instance of usage. This first example can be found at \docs.}
    \label{fig:Diagram}
\end{figure}

The overall structure of the \bd\ package is outlined in Figure \ref{fig:Diagram}. The package defines three Python classes:
\begin{itemize}
    \item \texttt{BubbleConfig} describes the background field. To initialise an object of this class requires a scalar potential $V(\phi)$, the false vacuum $\phi_\text{F}$, a bubble profile $\phi_\text{b}(r)$, and the dimension $d$.
    \item \texttt{ParticleConfig} describes a fluctuating particle. It is initialised by the function $W(\phi)$, the spin of the particle $s$, the number of its internal flavour or colour degrees of freedom $n$, and a flag denoting the type of zero modes present.
    \item \texttt{BubbleDeterminant} computes the functional determinant. It is initialised by one \texttt{BubbleConfig} instance and a list of \texttt{ParticleConfig} instances.
\end{itemize}
For more details, see the documentation and examples.

Once an instance of the class \texttt{BubbleDeterminant} is initialised, the most important class method is
\begin{align}
    &\texttt{findDeterminant()} = \sum_{i} \bigg[
    \frac{\text{dof}(d,s_i,n_i)}{2}
    \log\left|\frac{\det {'} (-\nabla^2 + W_i(r))}{\det(-\nabla^2 + W_i(\infty))}\right|
    - \log \mathcal{J}_i\mathcal{V}_i
    \bigg],
\end{align}
which computes a functional determinant in its totality, regularised in \MSbar, and with zero mode factors included where necessary. In this convention, the result is an additive correction to the action. Here the index $i$ runs over the \texttt{ParticleConfig} list. For scalar fields the factor $\text{dof}(d,0,n_i) = n_i$, and for vector fields $\text{dof}(d,1,n_i) = (d-1)n_i$ (see Appendix \ref{app:Vector}). Note that while Jacobian factors are included in all cases where zero modes exist, we do not include the (infinite) volume factor for the zero modes corresponding to translations. As a consequence, $\exp(-\texttt{findDeterminant()})$ has units of $\text{mass}^{d}$ for the Higgs determinant.

For classically conformal models the size of the bounce needs to be integrated over to obtain the rate. In \bd\ this integration does not change the mass-dimension of the rate. See the discussion in Appendix \ref{app:ConformalVolume}.

For thermal transitions, one should pass the keyword \texttt{thermal=True}. In this case, the dynamical prefactor term $- \log A_\text{dyn}$ is added to the output, given in terms of the negative eigenvalue and assuming zero damping coefficient.

The \texttt{BubbleDeterminant} class also contains a number of ancillary methods for computing different specific parts of the prefactor, such as the negative eigenvalue. Figure \ref{fig:Diagram} shows a component diagram of the structure of the package, and how it can be used in conjunction with \ct\ to compute the vacuum-decay rate for our introductory example.

\subsection{Application of the Gelfand-Yaglom method} \label{sec:GYNumerics}
From the Gelfand-Yaglom theorem, the determinant for a given orbital quantum number $l$ is given by the ratio $\psi_{b}^l(r)/\psi_{F}^l(r)$ at $r=\infty$. While this ratio is finite, both $\psi_{b}^l(r)$ and $\psi_{F}^l(r)$ grow exponentially at large $r$, which can lead to overflow for floating-point numbers. To avoid this, one can work directly with the ratio,%
\footnote{In fact, even $T_l(r)$ can get exponentially large for the fluctuations of heavy fields, in which case one can instead work with, $F_l(r) \equiv \log T_l (r)$. In the code $F_l$ is directly calculated for all $l\geq 2$, with the exception of the massless case, which is dicusssed in Appendix \ref{app:massless_algorithm}.}
\begin{align}\label{eq:TlFunc}
    T_l(r) \equiv \frac{\psi_{b}^l(r)}{\psi_{F}^l(r)}.
\end{align}
This function solves the following initial value problem
\begin{align}
    \left[-\partial^2 - U_l(r) \partial + \Delta W(r) \right]T_l(r) &= 0,
\end{align}
with $T_l(0) = 1$, $\partial T_l(0) = 0$, and where we have defined
\begin{align}\label{eq:UFunction}
    U_l(r) & \equiv \frac{d + 2l - 1}{r} + 2 \mF \frac{I_{l+\frac{d}{2}}(\mF r)}{I_{l+\frac{d}{2}-1}(\mF r)},
\end{align}
in terms of modified Bessel functions $I$.
We integrate the first step of the initial value problem, from 0 to $\delta r$, by using a Taylor expansion around the origin, and making use of the initial conditions; see Appendix \ref{app:ivp}.
We then proceed by using the fourth-order Runge-Kutta method to evolve from $\delta r$ to some $r_\text{max}$, the largest radius at which the bounce profile is given.

Errors on $T_l(\infty)$ arise from two main sources: discretisation errors due to non-zero radial steps $\delta r>0$, and the error due to non-infinite $r_\text{max}<\infty$. The former error scales as $(\delta r)^4$ as long as the bubble profile is known to this accuracy.%
\footnote{Bubble profiles computed with \ct\ are $(\delta r)^4$ accurate.}
The error due to the finite maximum radius is smaller than any power of $1/r_\text{max}$, for $\mF >0$, and for sufficiently large $r_\text{max}$, and we estimate it from the value of the derivative at the final step, $\partial T_l(r_\text{max})$. The total error on $T_l(\infty)$ is simply estimated as the larger of these two sources of error.

For the massless case $\mF = 0$, the solution converges relatively slowly as $r\to\infty$ so additional methods are adopted to accelerate convergence; see Appendix \ref{app:massless_algorithm}.

\subsection{Extrapolating the sum over orbital quantum number} \label{sec:ExtrapolationNumerics}
After extracting the $l=0$ and 1 modes, which need separate consideration, the logarithm of the complete functional determinant involves a sum over $l$ from 2 to $\infty$. The large $l$ divergences of this sum cancel against the one-loop counterterms of equation~\eqref{eq:CountertermAction}. For a single scalar field, the renormalised sum is thus
\begin{align}
   \frac{1}{2} \left[\sum_{l=2}^\infty \deg(d;l)\log T_l\right] + S_\text{ct} = \text{finite}.
\end{align}
To avoid divergences in intermediate computations, and to speed up convergence, we add and subtract a WKB approximation of $T_l$:
\begin{align} \label{eq:finite_sum}
    &\frac{1}{2}\sum_{l=2}^{l_\text{max}} \underbrace{\deg(d;l)\left(\log T_l - \log T_l^\text{(WKB)}\right)}_\text{finite and converges faster} +\underbrace{ \frac{1}{2}\sum_{l=2}^\infty \deg(d;l) \log T_l^\text{(WKB)}+S_\text{ct}}_\text{finite and known}
    \nonumber
    \\&\quad
    + \frac{1}{2}\underbrace{\sum_{l=l_\text{max}+1}^\infty \deg(d;l)\left(\log T_l - \log T_l^\text{(WKB)} \right)}_{\approx 0}.
\end{align}

The WKB factors $\log T_l^\text{(WKB)}$ are individually finite. Including them in the summand cancels the large $l$ dependence of $\log T_l$ up to $o(1/l)$, which ensures that the sum converges.
In practice, this sum is truncated at some large value $l_\text{max} \gg 1$, with the residual scaling as an inverse power of $l_\text{max}$, and the term labeled $\approx 0$ in equation \eqref{eq:finite_sum} is dropped.

To accelerate the convergence of this sum we utilise sequence acceleration, adopting two different methods: the epsilon algorithm \cite{graves2000epsilon} (which implements an iterated Shanks transformation \cite{bender1999advanced}), and a fit extrapolation, based on fitting a polynomial in $1/l$. For the latter, the order of the polynomial is chosen to minimise $\chi^2$ per degree of freedom. Typically the epsilon extrapolation performs better than the fit extrapolation when $l_\text{max}$ is relatively small, but performs worse for larger $l_\text{max}$ as numerical errors build up. We choose the final extrapolated result as that with smaller estimated error.

\begin{figure}[t]
    \centering
    \includegraphics[width=0.6\textwidth]{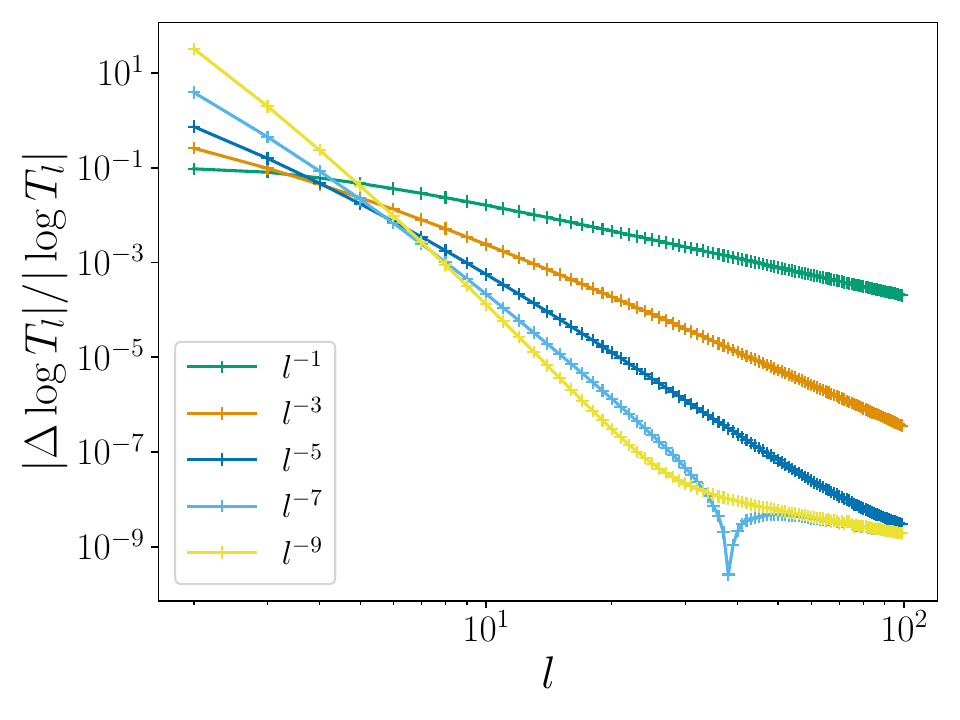}
    \caption{Convergence of the WKB approximation at large $l$. Shown is the relative difference between the complete $\log T_l$, calculated using the Gelfand-Yaglom theorem, and a number of WKB approximations to it. Higher order WKB approximations are used to accelerate the convergence of sums over $l$. The specific model is a real scalar with potential given by Eq.~\eqref{eq:V4Scaled}, at $\alpha=0.5$ and in $d=3$. The data for this plot is produced by the example \texttt{wkb.py}.}
    \label{fig:WKB}
\end{figure}

The terms $\log T_l^\text{(WKB)}$ are computed within the WKB approximation, the expressions for which are collected in Appendix~\ref{app:WKB}. Higher order terms of the WKB expansion are suppressed by higher powers of $1/l$, so that one need only compute a finite number of WKB terms to attain a given rate of convergence in $1/l$. By carrying out this approach to higher and higher orders, one can achieve much faster convergence of the sum, as demonstrated in Figure \ref{fig:WKB}.

Note that in higher dimensions the sum over $l$ becomes increasingly ultraviolet divergent, so higher orders of the WKB expansion are needed to renormalise the sum and to accelerate convergence. This is because the degeneracy factor grows faster in higher dimensions, as $\deg(d;l)\sim \bar{l}^{d-2}$ where $\overline{l}=l+\frac{d}{2}-1$, while the expansion for $\log T_l$ takes the same form in all dimensions,
\begin{align}
\log T_l \underbrace{=}_{l\gg \mF R}
\frac{\log T_l^{(1)}}{\overline{l}}
+\frac{\log T_l^{(3)}}{\overline{l}^3}
+\frac{\log T_l^{(5)}}{\overline{l}^5}
+\frac{\log T_l^{(7)}}{\overline{l}^7}
+\ldots
\end{align}
where $R$ is the radius of the critical bubble. When subtracting terms with the WKB approximation we therefore need to know (at least) the first $\lfloor\frac{d}{2}\rfloor$ terms to cancel all divergences.

An important numerical consideration is that one must compute each $\log T_l$ to higher accuracy in higher dimensions, as the necessary cancellations are more delicate. For example in $d=4$ and choosing $l_\text{max}\sim 25$, we need to know the numerically obtained $\log T_l$ to a relative accuracy of $l_\text{max}/l_\text{max}^3\sim 10^{-3}$. For $d=6$ we also need the next term, so we need to know $\log T_l$ to $l_\text{max}/l_\text{max}^5\sim 10^{-6}$.

In \bd\ we utilise the WKB approximation up to and including $\log T_l^{(9)}$; see Appendix \ref{app:WKB}. We do so for all dimensions, thereby ensuring a relatively rapidly converging sum in lower dimensions. Conversely, to achieve a fixed accuracy requires larger $l_\text{max}$ in higher dimensions.

\subsection{Finding the asymptotic behavior} \label{sec:PhiInfinityNumerics}
Here, we will discuss the implemented numerical method for finding \logphiinf, which is defined by
\begin{equation}\label{eq:asymptoticTailBehavior}
    \phi(r)\overset{r\to\infty}{\longrightarrow}\phi_{\text{F}}+\phi_\infty\,\text{K}(d/2-1,\mF r)\, \left(\frac{\mF}{r}\right)^{d/2-1}\,.
\end{equation}
Once we have determined \logphiinf, it is straightforward to deal with zero modes~\cite{Coleman:1978ae, Dunne:2006ct}. See for example equation~\eqref{eq:packTransZeroModes}.

An estimate for \logphiinf\ can be found by fitting the numerical bounce to equation \eqref{eq:asymptoticTailBehavior} for some sufficiently large radius.

Still, this must be done carefully since a numerically obtained profile will not exactly follow the asymptotic behaviour in equation \eqref{eq:asymptoticTailBehavior}. Large deviations may happen for example when a shooting algorithm stops as the bounce tail crosses the metastable vacuum. As the very tail end of a numerical bounce has little effect on the corresponding action, its accuracy is often not a high priority.

\begin{figure}[t]
    \centering
    \includegraphics[width=1.0\textwidth]{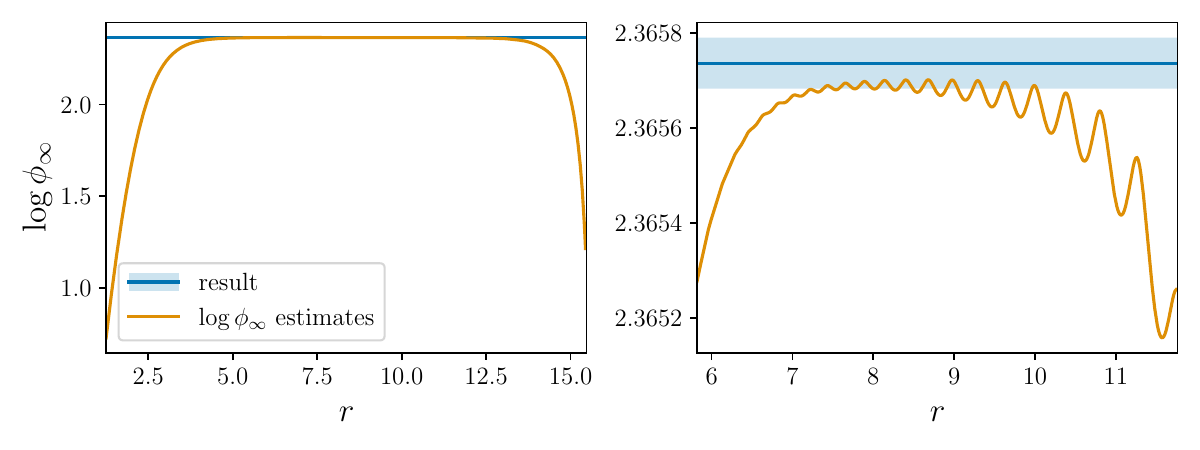}
    \caption{Estimates for \logphiinf\ by directly fitting the asymptotic tail behavior, Equation~\eqref{eq:asymptoticTailBehavior}, to points on the numerical critical bubble solution for the potential in Equation~\eqref{eq:singularPotential}, and the result from \bd\ with uncertainties. The right panel zooms in on the region of the left pane in which the estimate is nearly constant.}
    \label{fig:lnPhiInfRadPlot}
\end{figure}

The need for caution is illustrated in Figure~\ref{fig:lnPhiInfRadPlot}: The orange line is obtained by directly estimating \logphiinf\ at different radii. The bubble is produced with \ct\ in a model in three dimensions with scalar potential,
\begin{equation}\label{eq:singularPotential}
    V(\phi) = \frac{\mF^2}{2}\phi^2+\frac{\lambda}{4!}\phi^4\log\phi\,,\quad \mF^2=1\,,\quad\lambda=100\,.
\end{equation}
Note how the estimate begins to diverge at large radial distances, $r>10$. Also, there are oscillations, visible on the right pane, which follow from the use of cubic splines in \ct. The peaks of the oscillations correspond to the best estimates as they are the nodes of the splines.

To evade these problems at the tail-end, we have constructed an algorithm that is robust even when the tails are inaccurate. In the following, we focus on the massive case, $\mF >0$. An example result from the algorithm is plotted in blue in Figure~\ref{fig:lnPhiInfRadPlot}, along with the direct estimates.

The algorithm takes two user-given parameters: \tail, which determines how close to the end of the profile to fit, and \lpit, which gives an estimate of the relative uncertainty on the tail of the profile.

The main algorithm then finds \logphiinf\ by extrapolating from the chosen \tail; see Figure~\ref{fig:lnPhiInfLinear} and the surrounding discussion. If there are no such suitable tail points, if for example the bounce profile is too short, the code resorts to a fail-safe algorithm, which instead performs the fit at $r=\sqrt{r_{1/2}r_\text{max}}$, where $\phi_\text{b}(r_{1/2})=(\phi_\text{F}+\phi_\text{T})/2$ and $r_\text{max}$ is the largest radius. The error is then estimated by comparison to fits at radial distances $(r_{1/2}^2r_\text{max})^{1/3}$ and $(r_{1/2}r_\text{max}^2)^{1/3}$.

In case the parameter \lpit\ does not correctly reflect errors in the bubble profile, the more sophisticated main method is run a second time but with \lpit\ modified to give at most the error of the fail-safe method. Together with the first run of the main method and the fail-safe method, this gives three estimates for \logphiinf, of which that with the least error is returned.

We refer to Appendix \ref{app:phiInf} for a more detailed description of the fitting algorithm, and for a discussion of the massless case, $\mF = 0$.

\subsection{Computing the negative eigenvalue} \label{sec:NegativeEigenvalueNumerics}
For thermal nucleation, the decay rate factorises into a product of statistical and dynamical parts, as is shown in equation \eqref{eq:ThermalRate}. While the statistical part can be computed through application of the Gelfand-Yaglom method, this is not so for the dynamical part. In the latter, the negative eigenvalue of the Higgs operator $\mathcal{O}_H(\phi_\text{b})$ appears in combination with the real-time damping rate. For computation of the negative eigenvalue, \bd\ provides the function \texttt{findNegativeEigenvalue()}.

The negative eigenvalue, $\lambda_-<0$, is defined by the following eigenvalue problem,
\begin{equation}\label{eq:diffEigValProblem}
    \left(-\partial^2-\frac{d-1}{r}\partial+V''(\phi_\text{b})\right)f(r)=\lambda_- f(r)\,,\quad r\in (0,\infty)\,,
\end{equation}
where $\partial f(r)\to 0$ as $r\to 0^+$ and $f\to 0$ as $r\to\infty$. Note, that here we have used the information that the negative eigenmode is spherically symmetric, $l=0$.

The eigenvalue problem can be approximated as a discrete matrix equation,
\begin{equation}\label{eq:discretizedEigenvalueProblem}
    M_{ij}f_j=\lambda_- f_i\,,\quad i,j\in \{1,\dots,N\}\,.
\end{equation}
Here, the matrix $M$ is a discretization of the linear differential operator in Eq.~\eqref{eq:diffEigValProblem}, such that all of the rows (indices $i$) correspond to individual locations of $r$, and $N$ is the number of discrete points in the numerical critical bubble.

The code obtains a numerical estimate of the negative eigenvalue by passing $M$ to \texttt{scipy.sparse.linalg.eigs}.
This estimate is then improved by an extrapolation: The implemented derivatives, discussed below, are
accurate to such an order
that the errors in $\lambda_-$ decrease as $N^{-4}$. Obtaining the estimates with one half and one third of the points allows for an extrapolation to $N\to \infty$ and also an error estimation.

The boundary condition at infinite radius, $r\to\infty$, cannot be implemented accurately in the discretized version. To estimate the size of the corresponding error, in the code there are two alternative boundary conditions implemented at the maximal numerical radius, $r_{\text{max}}$:
\begin{equation}
    \partial f(r_{\text{max}})=0\quad\text{and}\quad f(r_{\text{max}})=0\,.
\end{equation}
In the limit $r_{\text{max}}\to\infty$, these both reduce to the correct boundary condition for the negative eigenmode. Deviations are exponentially small for a large enough maximal radius, $\propto \exp(-2\sqrt{\mF^2-\lambda_-}\,r_{\text{max}})$.

Values with errors are computed for each of the two boundary conditions, extrapolated to $N\to\infty$. The final result and error are given as the midpoint and extent of the combined uncertainty range.

We refer to Appendix \ref{app:NegativeEig} for more details.

\subsection{Requirements on the input bounce profile} \label{sec:BounceNumerics}
Before computing the functional determinant, one must solve for the bounce. While \bd\ does not provide this functionality, there are a number of other packages which are built for this purpose: \ct\ in Python \cite{Wainwright:2011kj}, \texttt{AnyBubble} and \texttt{FindBounce} in Mathematica \cite{Masoumi:2016wot, Guada:2020xnz}, and \texttt{BubbleProfiler} and \texttt{SimpleBounce} in C++ \cite{Athron:2019nbd, Sato:2019wpo}.

As a word of caution, for computation of the functional determinant, the bounce profile has to be known rather accurately, and out to rather large $r$. Unlike the bounce action, the functional determinant depends strongly on the large $r$ asymptotics of the bounce, and further fluctuations with high orbital quantum numbers probe the short scale structure. With \ct\, we have found that using tolerance parameters \texttt{xtol=1e-9} and \texttt{phitol=1e-9} typically yields an accurate enough bounce profile for computing the functional determinant reliably. Note that this is significantly more accurate than \ct's default values.

\section{Tests} \label{sec:Tests}

\subsection{Comparisons to literature}

We have carried out extensive tests of the \bd\ code against the literature, comparing against all the explicit results we could find. We have generally found agreement within quoted errors.

\newtext{A number of authors have numerically computed the one-loop vacuum decay rate for a single real scalar field in a generic quartic tree-level potential \cite{Baacke:2003uw, Dunne:2005rt, Ekstedt:2021kyx},}
\begin{align} \label{eq:V4}
&V_4(\phi)=\frac{1}{2}m^2\phi^2-\frac{1}{2}\eta \phi^3+\frac{1}{8}\lambda \phi^4.
\end{align}
\newtext{Here the subscript in $V_4$ refers to the highest power of $\phi$ present in the potential.} By using the result in Appendix \ref{app:Potentials:Phi4} the action can be brought to the form
\begin{align} \label{eq:V4Scaled}
S[\phi]\rightarrow \beta S[\phi],\quad
V_4(\phi)\rightarrow \frac{1}{2}\phi^2-\frac{1}{2}\phi^3+\frac{1}{8}\alpha \phi^4,
\end{align}
where $\alpha=\lambda m^2\eta^{-2}$ and $\beta=m^{6-d}\eta^{-2}$. This dimensionless form is quite useful since the determinant is\te up to a $-\frac{d}{2}\log \beta$ factor\te only a function of $\alpha$. \newtext{Note that a term linear in $\phi$ can be removed by a shift $\phi \to \phi + \text{const}.$}

\newtext{Refs.~\cite{Baacke:2003uw, Dunne:2005rt} have carried out the computation in this potential for $d=4$.} We find agreement to better than 1\% with the tabulated results of Ref.~\cite{Baacke:2003uw} for the full of parameters studied, $\alpha \in [0.01, 0.99]$. With Ref.~\cite{Dunne:2005rt}, we find good agreement with their figures~5 and 6, though we disagree on their figure~7.

We have also compared the result from \bd\ to exact analytical results for classically scaleless potentials; for example the four-dimensional potential $V(\phi)=-\frac{1}{4} \lambda\phi^4$~\cite{Fubini:1976jm, Lipatov:1976ny}. Previous results exist in four dimensions~\cite{Andreassen:2017rzq, Isidori:2001bm, Ivanov:2022osf}, and as an additional crosscheck we have also derived analytical results for three and six dimensions; to our knowledge these results don't exist in the literature, but the derivation is identical to the four-dimensional case and won't be repeated here\footnote{\newtext{The four-dimensional result can also be used to describe more general models, see~\cite{Chigusa:2018uuj}.}}. A summary of the results can be found in table \ref{table:ConformalBounces}.
\begin{table}[t]
\begin{center}
    \begin{tabular*}{0.88\textwidth}{@{\extracolsep{\fill} } c | c | c | c| c }
    \hline
  \centering  $d$ & \centering $V(\phi)$ & \centering $\phi_\text{b}$ & \centering $S_0$& $S_1$ \\ \hline
    $3$ & -$\frac{1}{6}\lambda \phi^6$ &$\left(\frac{3}{\lambda}\right)^{1/4}\sqrt{\frac{R}{r^2+R^2}}$  &$\frac{\pi^2}{2}\sqrt{\frac{3}{\lambda}}$&$3\log R + \log\lambda + 4.41321$ \\ \hline
    $4$ & -$\frac{1}{4}\lambda \phi^4$ &$\sqrt{\frac{8}{\lambda}}\frac{R}{r^2+R^2}$  &$\frac{8 \pi^2}{3 \lambda}$&$4\log R + \frac{5}{2}\log \lambda - 3\log\mu R -0.991929$ \\ \hline
    $6$ & -$\frac{1}{3}\lambda \phi^3$ &$\frac{24}{\lambda} \frac{R^2}{\left(r^2+R^2\right)^2}$  &$\frac{192 \pi^3}{\lambda^2}$&$6\log R + 7\log\lambda - \frac{18}{5}\log\mu R - 16.1573$ \\ \hline
    \end{tabular*}
    \caption{\label{table:ConformalBounces} Results for scale-invariant models with unbounded potentials ($\lambda>0$) in $d=3$, 4 and 6. \newtext{Here $S_0$ and $S_1$ denote the tree-level and one-loop effective action respectively, i.e.~$S_0=B$ and $S_1=-\log A$ in equation \eqref{eq:intro_saddlepoint}.} The numerical factors in $S_1$, correspond to various combinations of Riemann zeta functions and Euler-Mascheroni constants. The full rate per unit volume is given by $\Gamma = \int d(\log R) e^{-S[R]}$.}
\end{center}
\end{table}

\begin{figure}[t]
    \centering
    \begin{subfigure}{.5\linewidth}
\centering
\includegraphics[width=1\textwidth]{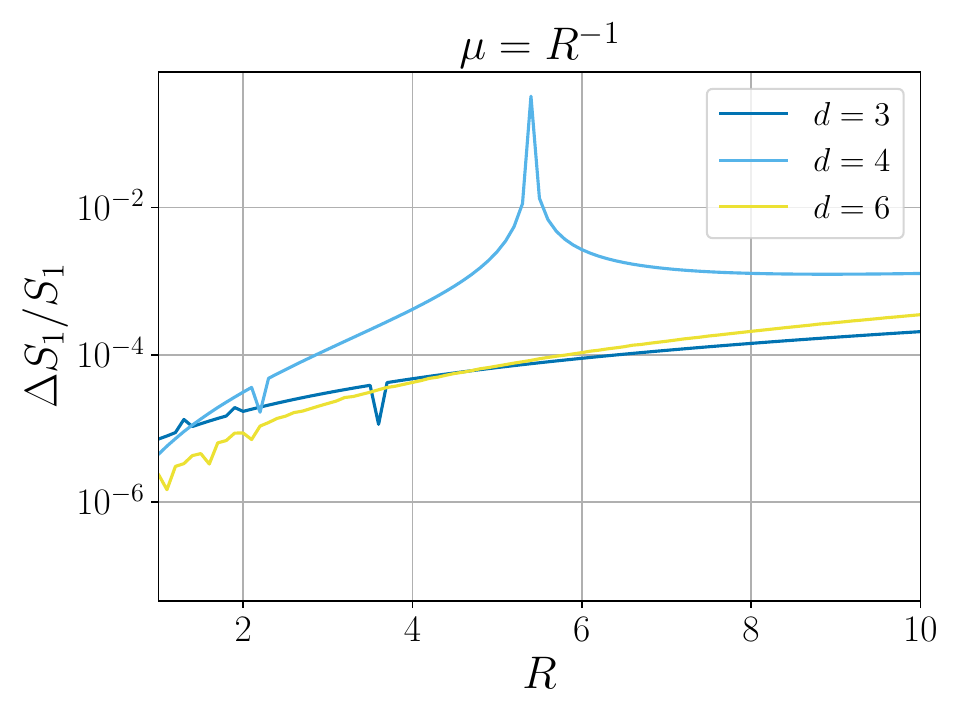}
\caption{}
\label{fig:Scaleless}
\end{subfigure}%
\begin{subfigure}{.5\linewidth}
\centering
\includegraphics[width=1\textwidth]{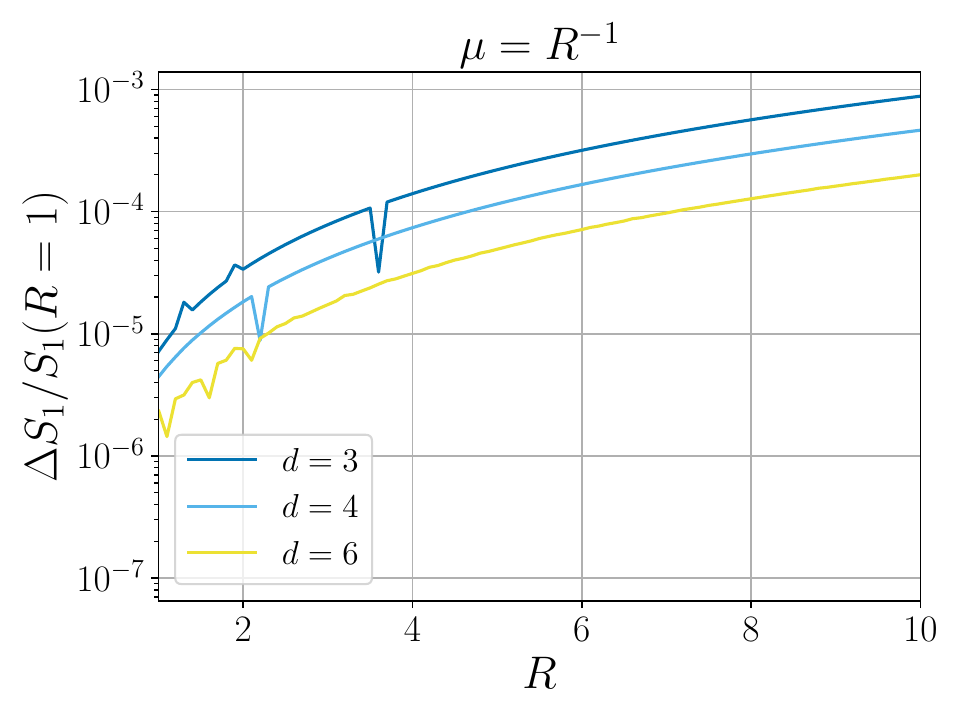}
\caption{}
\label{fig:ScalelessNorm}
\end{subfigure}
    \caption{The Higgs determinant, for given $R$, compared between \bd\ and the analytical results for the scaleless potentials given in Table \ref{table:ConformalBounces}. The left-hand plot shows the relative error as a function of $R$. Note that larger  errors in $d=4$ arise because $S_1(R)\approx 0$; this can be seen clearly in the right-hand plot where the relative difference is instead normalized by $S_1(R=1)$. All lines use the renormalization-scale $\mu=R^{-1}$ and $\lambda=0.1$. The analytical bounces, given in table \ref{table:ConformalBounces}, are supplied to \bd, which then calculates the one-loop action using default settings.
    }
	\label{fig:Unbounded}
\end{figure}

Figure \ref{fig:Unbounded} compares the results from \bd\ to the exact results for scaleless potentials given in table~\ref{table:ConformalBounces}. \newtext{There we have introduced $S_1$ to denote the one-loop correction to the effective action, i.e.~minus the logarithm of the functional determinant prefactor.} In all cases we find agreement at better than 1\%, excepting a very narrow range where $S_1$ goes through zero in $d=4$, and in most cases the agreement is better than 0.1\%. For this model, exact results are also available for the decomposition of the determinant into orbital quantum numbers $l$, and for fixed $l$ we find even better relative agreement, as shown in the example file \texttt{unbounded.py}.

In $d=3$, Ref.~\cite{Ekstedt:2021kyx} numerically computed the Higgs, Goldstone and vector field determinants for the same potential, given by equation~\eqref{eq:V4}, as well as for 
\begin{align} \label{eq:V6}
&V_6(\phi)=\frac{1}{2}m^2\phi^2-\frac{1}{4}|\lambda|\phi^4+\frac{1}{32}c_6 \phi^6.
\end{align}
which, after using the result in Appendix \ref{app:Potentials:Phi6}, becomes
\begin{align} \label{eq:V6Scaled}
S[\phi]\rightarrow \beta S[\phi],\quad
V_6(\phi)=\frac{1}{2}\phi^2-\frac{1}{4}\phi^4+\frac{1}{32}\alpha \phi^6.
\end{align}
with $\alpha=c_6 m^2|\lambda|^{-2}$ and $\beta=m^{4-d}|\lambda|^{-1}$.

For the $V_4$ potential in $d=3$ we find agreement with the fit functions of Ref.~\cite{Ekstedt:2021kyx} to the 1\% level, and for the $V_6$ potential we find agreement to the 1\% to 5\% level, except for where $S_1$ goes through zero, near $\alpha\approx 0.4$. This agreement matches the expected accuracy of the fits.

\subsection{Thin-wall limit}

In the thin-wall limit there are a number of analytic results for the one-loop vacuum decay rate, in $d=2$ \cite{Munster:2000qt, Munster:2003an}, $d=3$ \cite{Munster:1999hr, Ivanov:2022osf} and $d=4$ \cite{Konoplich:1987yd, Ivanov:2022osf}, all of which use the potential of equation~\eqref{eq:V4Scaled} up to trivial scalings and shifts. We also used a method similar to that in \cite{Ivanov:2022osf} to derive new analytical results for $d=5$, 6 and 7.

\begin{figure}[t]
    \centering
    \includegraphics[width=0.85\textwidth]{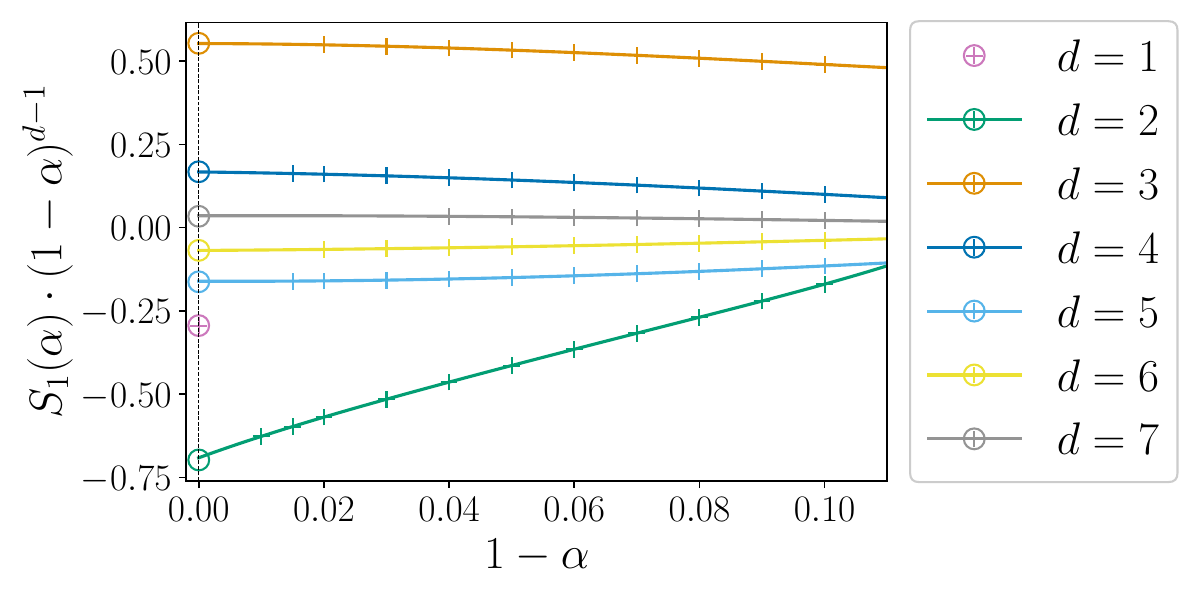}
    \caption{The approach to the thin-wall limit in $d=2, 3, \dots 7$. Unfilled circles show the analytic results valid in the limit $\alpha\to 1$. Crosses show data computed using \bd, and the lines show cubic fits to this data. We also show the nearest equivalent for $d=1$, the one-loop correction to the energy eigenvalue splitting. The data for this plot can be generated using our example scripts \texttt{thinwall.py} and \texttt{kink.py}.
    }
    \label{fig:Thinwall}
\end{figure}

With our potential conventions, and in the \MSbar\ renormalisation scheme with the \MSbar\ scale equal to the mass, these results read
\begin{align}
S_1\newtext{(\alpha)} = \frac{1}{(1-\alpha)^{d - 1}}\times
\begin{cases}
-1 + \frac{\pi}{6\sqrt{3}}, & d=2 \\
\frac{10}{27} + \frac{\log 3}{6}, & d=3 \\
\frac{9}{32} - \frac{\pi}{16\sqrt{3}}, & d=4 \\
-\frac{296}{3645}-\frac{2 \log 3}{27}, & d=5 \\
 -\frac{1376251}{11197440}+\frac{625 \pi }{20736 \sqrt{3}},  & d=6 \\
 -\frac{1}{140} + \frac{3\log 3}{80}, & d=7 \\
\end{cases}
\end{align}
In Figure \ref{fig:Thinwall} we show how our numerical results approach the analytic thin-wall results in the approach to the thin-wall limit. Fitting a cubic function of $\alpha$ to the data yields extrapolations for $\alpha \to 1$ which agree with the expected analytical values to better than 1\% accuracy for dimensions $d=2$, 3, \dots, 6.

For $d = 7$, the increased sensitivity to shorter distance fluctuations requires including higher orbital quantum numbers (see Section \ref{sec:ExtrapolationNumerics}). This in turn prevents us from reaching $1-\alpha\lesssim 0.05$ while keeping numerical errors under control, and hence significantly worsens the accuracy of the $\alpha \to 1$ extrapolation in $d=7$, which agrees to only about 10\% with the analytic result.

Note that the thin-wall limit is a particularly difficult regime numerically, as to resolve the scale hierarchy $m_F R \gg 1$ requires very high orders in the sum over orbital quantum number $l$, which leads to accumulation of numerical errors. As a consequence, we expect \bd\ to perform better, and to achieve higher accuracy away from the thin-wall limit, where analytic results are lacking.

For $d=1$, there is one Euclidean time dimension \newtext{and} zero spatial dimensions, and hence we are treating tunnelling in quantum mechanics. At $\alpha=1$, the potential of equation~\eqref{eq:V4Scaled} corresponds to the double-well potential, which admits kink or soliton solutions \cite{Dashen:1974cj}.
These are quantum mechanical instantons. The functional determinant in a kink background gives the one-loop correction to the splitting between the even and odd energy eigenstates \cite{Coleman:1978ae, Zinn-Justin:1982aya}. Evaluating this analytically, one finds
\begin{align}
    S_1 = \frac{1}{2}\log\pi - \frac{5}{4}\log 2.
\end{align}
With default settings, \bd\ reproduces this one-loop result to a relative accuracy of 0.001\%. This is shown in Figure \ref{fig:Thinwall}. Similar agreement is found for the sine-Gordon theory, $V_\text{s-G}(\phi)=1-\cos(\phi)$, which also admits kink solutions in $d=1$ and for which $S_1 = \frac{1}{2}\log\pi - 2\log 2$.

\subsection{Derivative expansion}
\begin{figure}[t]
    \centering
    \includegraphics[width=0.7\textwidth]{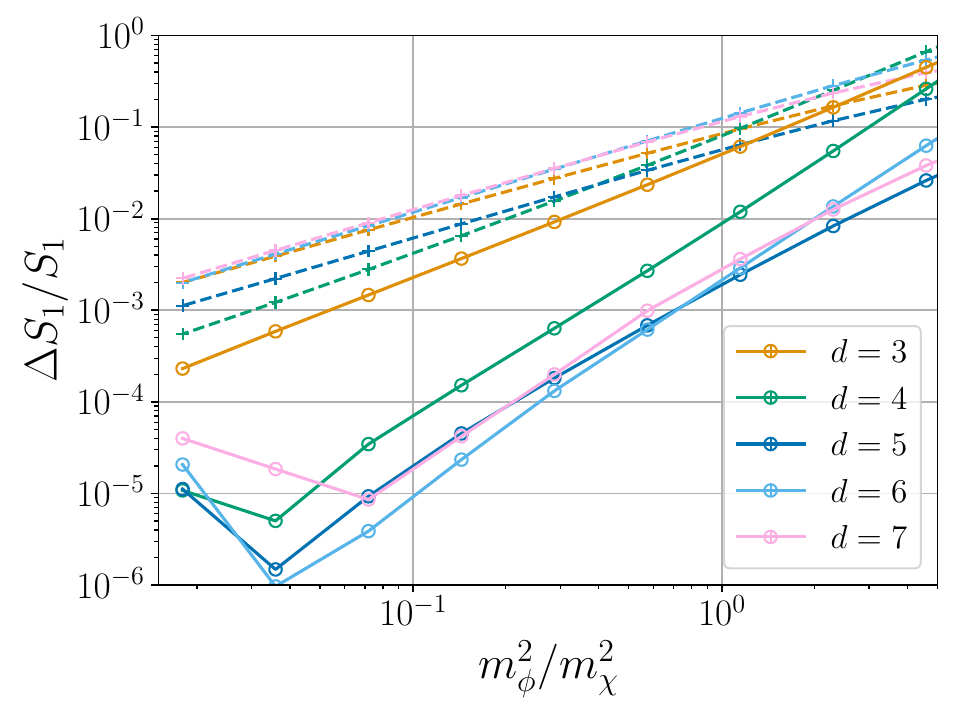}
    \caption{Deviations from the derivative expansion for the determinant of the $\chi$ field; see equation \eqref{eq:LagrangianDerivativeExpansion}. Here the value of $m_\phi^2/m_\chi^2$ is given in the true vacuum. Crosses connected by dashed lines show the LO derivative expansion, while unfilled circles connected by full lines include the NLO order corrections. For small $m_\phi^2/m_\chi^2$, convergence appears to arrest at about $10^{-5}$, especially in higher dimensions. This is because one must reach very large values of the angular quantum numbers $l$, and in so doing numerical errors accumulate (here on default tolerance settings).}
	\label{fig:DerivativeExpansion}
\end{figure}
For fluctuations of some scalar field $\chi$, much heavier than the background Higgs field $\phi$, a derivative expansion of the functional determinant may be possible. This is an expansion in powers of the ratio of masses squared, $m_\phi^2/m_\chi^2$. Consider, for example,
\begin{align}
    \mathscr{L}_\chi = \frac{1}{2}\nabla_\mu \chi \nabla_\mu \chi + \frac{1}{2}g^2\phi^2 \chi^2,
\end{align}
so that $m_\chi^2 = g^2 \phi^2$. For sufficiently large $g^2$, the functional determinant for $\chi$ should be well approximated by a derivative expansion. At leading order (LO) and next-to-leading order (NLO), the result takes the form
\begin{align} \label{eq:LagrangianDerivativeExpansion}
    S_1 \approx \int \mathrm{d}^d x \left[
            V_{1}(\phi)
            + \frac{1}{2}Z_{1}(\phi)\nabla_\mu \phi \nabla_\mu \phi
        \right],
\end{align}
where $V_1$ and $Z_1$ are the contribution to the effective potential and field normalisation for $\phi$ due to fluctuations of the $\chi$ field.
Expressions for $Z_{1}$ in arbitrary dimensions can be constructed from the results of Ref.~\cite{PhysRevD.46.1671}.

In Figure \ref{fig:DerivativeExpansion} we plot the relative difference between the full functional determinant and its LO and NLO approximations within the derivative expansion, for dimensions $d>2$. For $d=2$ the derivative expansion in this model is infrared divergent (as $r\rightarrow \infty$) at NLO, hence we do not include it.%
\footnote{Although not shown here, in $d=2$ we find agreement with the LO derivative expansion as expected.}
For $d=3$ the derivative expansion is infrared divergent at next-to-next-to-leading order (NNLO), and in fact there exists a term between NLO and NNLO which is invisible to the derivative expansion; see for example Ref.~\cite{Gould:2021ccf}. This explains why the slope of the NLO line in Figure \ref{fig:DerivativeExpansion} is approximately 3/2 and not 2.
In general as the coupling $g^2$ is increased, agreement with the derivative expansion improves, reaching 0.001\% at NLO in dimensions where the derivative expansion is under control. This provides a nontrivial test of \bd, which works as expected.

\subsection{Speed tests}
\begin{figure}[t]
    \centering
    \includegraphics[width=0.7\textwidth]{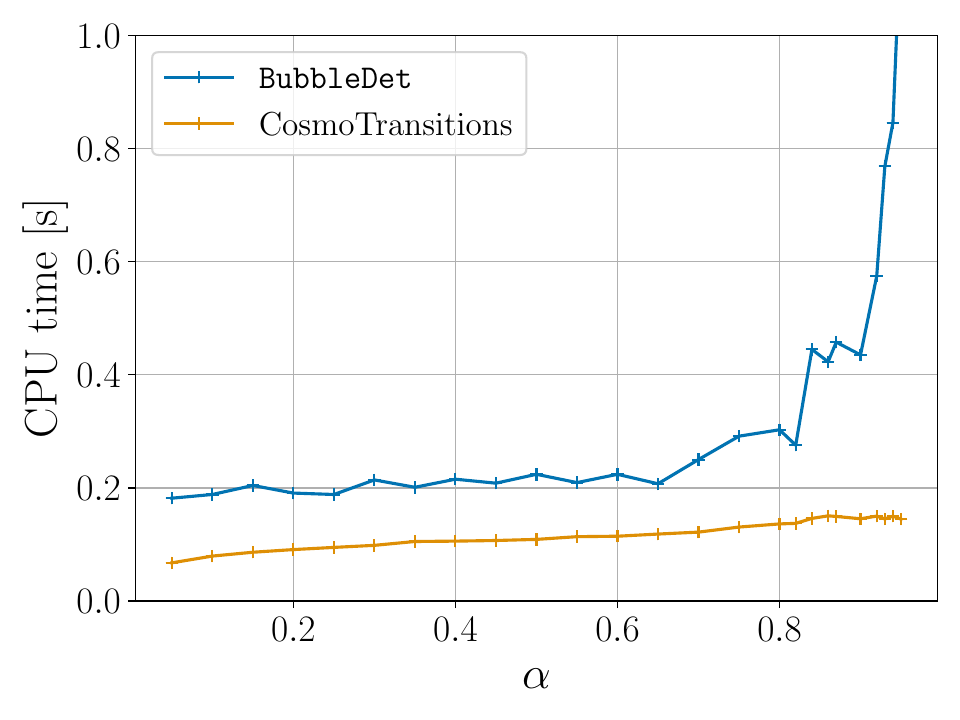}
    \caption{Time required to calculate the Higgs determinant for the potential given in equation~\eqref{eq:V4Scaled} with $d=3$. Computations were carried out on a laptop with a 3.2 GHz Apple M1 processor. Each data point represents the average of 300 runs. The required time to calculate the bounce action with \ct\ is shown in orange, and the required time for the determinant is shown in blue. \newtext{The latter increases in the approach to the thin-wall limit due to the necessity of reaching larger orbital quantum numbers, $l_\text{max} \gg \mF R$, where $R$ is the bubble radius.}}
	\label{fig:Timing}
\end{figure}

For speed comparisons we compare the time it takes to calculate the determinant relative to the time it takes for \ct\ to obtain the bounce solution. From Figure \ref{fig:Timing} we see that, on average, it only takes twice as long to find the bubble determinant as the bounce. \newtext{Note that our use of a sequence acceleration, combined with using more terms in the WKB approximation, is crucial.}

\section{Applications}

As argued in the introduction, functional determinants in quantum field theory exponentiate, and can thereby yield important corrections to nucleation or decay rates. Here we illustrate this, applying \bd\ to a range of scenarios.

\subsection{Thermal bubble nucleation and dimensional reduction}

For a simple example of a first-order phase transition, let us consider a Yukawa model describing a real scalar $\Phi$ interacting with a Dirac fermion $\Psi$. The model has the following Euclidean action (in $d=4$)
\begin{align}
    S_\text{Yukawa} =& \int \mathrm{d}^4x\ \bigg[
    \frac{1}{2}\left(\nabla_\mu \Phi\right)^2
    + s \Phi
    + \frac{1}{2} m^2 \Phi^2
    + \frac{1}{3!} g \Phi^3
    + \frac{1}{4!} \lambda \Phi^4 \nonumber \\
    &
    + \bar{\Psi}\left(\slashed{\partial}
    + m_\Psi\right)\Psi
    + y\Phi \bar{\Psi}\Psi 
    \bigg].
    \label{eq:yukawa}
\end{align}
The notation is standard and follows Ref.~\cite{Gould:2021ccf}. The thermal evolution of the long-wavelength modes of the field $\Phi$ are described by a dimensionally-reduced effective field theory (in $d=3$), which is
\begin{align}
	S_\text{EFT} =& \int \mathrm{d}^3x\ \bigg[
    \frac{1}{2}\left(\nabla_i \phi\right)^2
    + s_3 \phi
    + \frac{1}{2} m_3^2 \phi^2
    + \frac{1}{3!} g_3 \phi^3
    + \frac{1}{4!} \lambda_3 \phi^4
     \bigg].
\end{align}
where to leading order $\phi$ is equal to the zero Matsubara mode of $\Phi$ divided by $\sqrt{T}$, and the effective parameters are
\begin{align}
    s_3 &= \frac{1}{\sqrt{T}}\left[s + \frac{1}{24}(g+4ym_\Psi)\right], & g_3 &= g \sqrt{T}, \\
    m_3^2 &= m^2+\frac{1}{24}(\lambda+4y^2), & \lambda_3 &= \lambda T.
\end{align}
By shifting $\phi\to\phi+\text{const}$ and scaling, this model can be made to agree with the conventions of our $V_4$. However, here we work directly with the dimensionful parameters of the effective field theory.

As given in equation \eqref{eq:ThermalRate}, the thermal nucleation rate is the product of two terms. The statistical part $A_\text{stat}$ is the vacuum decay rate for the 3-dimensional EFT. For the dynamical part $A_\text{dyn}$ we neglect damping. The total decay rate is then
\begin{align}
    \Gamma_T = \frac{\sqrt{|\lambda_-|}}{2\pi}
    \left(\frac{S_\text{EFT}[\phi_\text{b}]}{2\pi}\right)^{3/2}
    \left\vert
    \frac{\det'\left(-\nabla^2+V_\text{EFT}''(\phi_\text{b})\right)}{\det\left(-\nabla^2+V_\text{EFT}''(\phi_\text{F})\right)}
    \right\vert^{-1/2}
e^{-S_\text{EFT}[\phi_\text{b}]+S_\text{EFT}[\phi_\text{F}]}.
\end{align}
The fluctuation determinant runs over the degrees of freedom of the EFT, i.e.~the 3-dimensional scalar field $\phi$. The fermion contributes to the nucleation rate through the temperature dependence of the effective parameters. Note that there is no factor of $1/T$ in the exponent, as factors of $T$ have been absorbed into the effective parameters and field.

\begin{figure}[t]
    \centering
    \includegraphics[width=0.6\textwidth]{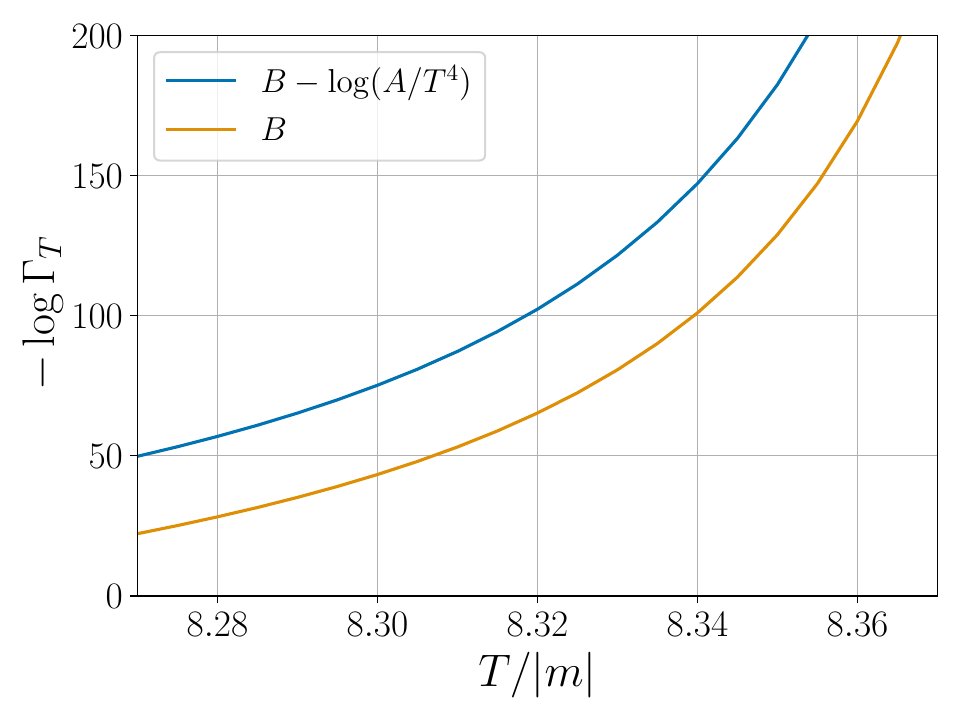}
    \caption{Thermal prefactor corrections for a Yukawa model; see equation.~\eqref{eq:yukawa}. Here we show the variation with temperature for one physical parameter point. The data for this plot can be generated using our example script \texttt{thermal.py}.}
    \label{fig:ThermalYukawa}
\end{figure}

In Figure~\eqref{fig:ThermalYukawa} we show the nucleation rate for an example thermal history of this model, with parameters $\{s,m^2,g,\lambda,m_\Psi,y\}=\{0,-1,0.3,0.1,-0.2,0.3\}$. At this parameter point the critical temperature is $T_\text{c}\approx 8.42 |m|$, and the one-loop correction to the jump in the order parameter $\Delta\langle\phi\rangle_\text{c}$ is about 8\% \cite{Gould:2021dzl}. However, the relative importance of loop corrections grows as the system supercools. Figure~\eqref{fig:ThermalYukawa} shows that the functional determinant can yield sizeable corrections to the rate, and that these eventually become nonperturbative in the approach to spinodal decomposition, at $T_\text{sp}\approx 8.19 |m|$.

\subsection{Gauge symmetry breaking}

Let us consider the Abelian Higgs theory in $d=4$, with Lagrangian given by equation \eqref{eq:AbelianHiggsLagrangian}. To simplify the analysis \cite{Ai:2020sru}, we include a $(\phi^*\phi)^3$ operator in the potential to generate a tree-level barrier between symmetric and broken phases; see equation \eqref{eq:V6}.

The full nucleation rate is then given by (see Appendix \ref{app:Vector})
\begin{align} \label{eq:RateSymmetryBreakingExample}
\Gamma \approx
\left(\frac{\det\mathcal{O}_{A_\mu}(\phi_\text{F})}{\det\mathcal{O}_{A_\mu}(\phi_\text{b})}\right)^{3/2}
\mathcal{V}_{G} \mathcal{J}_{G}
\sqrt{\frac{\det\mathcal{O}_{G}(\phi_\text{F})}{\det'\mathcal{O}_{G}(\phi_\text{b})}}
\mathcal{J}_{H}
\sqrt{\left|\frac{\det\mathcal{O}_{H}(\phi_\text{F})}{\det'\mathcal{O}_{H}(\phi_\text{b})}\right|}
    e^{-(S_0[\phi_\text{b}] - S_0[\phi_\text{F}])}.
\end{align}
We remind the reader that we have dropped off-diagonal gauge-Goldstone mixing.

\begin{figure}[t]
    \centering
    \includegraphics[width=0.6\textwidth]{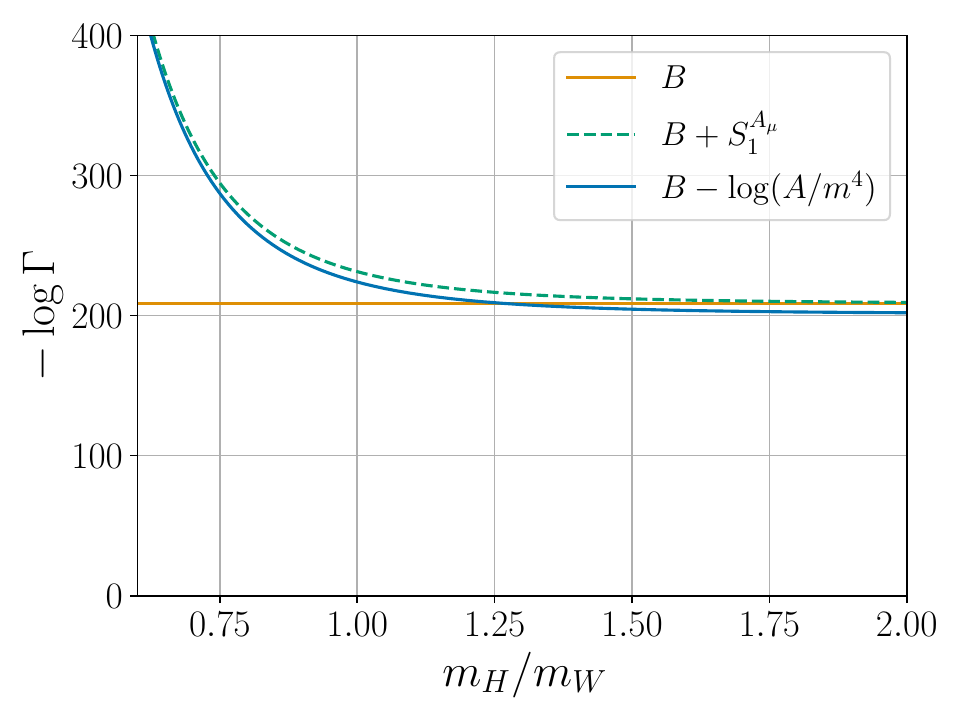}
    \caption{The vacuum decay rate, for spontaneous symmetry breaking, in the Abelian Higgs model. Here we have defined $S_1^{A_\mu}$ as the contribution to the 1-loop effective action from the gauge boson. As can be seen, this dominates over the scalar contributions for $m_H/m_W \lesssim 1$. The data for this plot is for fixed scalar parameters $m^2$, $\lambda$ and $c_6$, and varying gauge coupling $g^2$. It can be generated using our example script \texttt{symmetry\_breaking.py}.
    }
    \label{fig:SymmetryBreaking}
\end{figure}

Figure \ref{fig:SymmetryBreaking} shows the tree-level and one-loop corrections to the nucleation rate for the parameter point $\{m^2, \lambda, c_6\} = \{1, -0.15, 0.01\}$, varying the gauge coupling.

For mass ratios below about $m_H/m_W \lesssim 2/3$, or gauge couplings above $g^2\approx 1.3$, the functional determinant of the gauge field grows larger than the tree-level bounce action. For such large gauge couplings, the gauge determinant is exponentially enhanced by the ratio of gauge to scalar masses, or equivalently by $g^2/\lambda \gg 1$. In this case the broken perturbative expansion can be repaired by integrating out the gauge field before solving the bounce equation \cite{Weinberg:1992ds}.

\subsection{Analogue false vacuum decay in \texorpdfstring{$d=1+1$}{d=1+1}}

In studies of analogue false vacuum decay, a nucleating scalar field with an effective relativistic energy-momentum relation can arise as an angular degree of freedom in cold atom setups.%
\footnote{For a recent experimental observation of false vacuum decay in an alternative setup, see Ref.~\cite{Zenesini:2023afv}.}
In this case a trigonometric potential arises \cite{Fialko:2014xba, Billam:2020xna, Braden:2022odm}
\begin{align}\label{eq:CosinePot}
V(\phi)=m^2 \overline{v}^2\left[\cos(\phi/\overline{v})-1+\frac{\lambda^2}{2} \sin^2(\phi/\overline{v})\right].
\end{align}

The authors of Ref.~\cite{Braden:2022odm} investigated the effects of renormalisation in classical stochastic lattice simulations of vaccuum decay, studying parameter points $\lambda=2$, $\bar{v}^2\in [2, 10]$, and we presume $m^2=1$. They suggested a definition for a lattice effective potential aiming to capture ``the renormalization effects present in the fluctuation determinant'', though the latter was not directly computed. Using the lattice effective potential led to a roughly $e^{30}$ increase of the nucleation rate relative to tree-level, which was almost independent of $\bar{v}$ over the range studied.

For the parameter points studied in Ref.~\cite{Braden:2022odm}, we have used \bd\ to further investigate their proposal. In support, we indeed find that the functional determinant leads to an increase in the nucleation rate which varies little with $\bar{v}$ over the range studied. However, we find the magnitude of the increase to be significantly smaller, being only a factor of $\sim e^7$. This calculation can be found in our example script \texttt{analogue.py}.
A full comparison would require matching regularisation schemes. However, recent work has cast doubt on the continuum limit of these classical stochastic lattice simulations \cite{Hertzberg:2020tqa, Tranberg:2022noe}.

\subsection{Effects of potential shape}

To better understand the effects of potential shape on the functional determinant, it is useful to work with the dimensionless potentials introduced above, which have been put into a common form. The potentials are linear in $\alpha$, and $\alpha\rightarrow 0$ corresponds to the thick-wall limit, while $\alpha\rightarrow 1$ corresponds to the thin-wall limit. In addition, the determinants only depend nontrivially on $\alpha$, and not $\beta$.

In addition to the polynomial $V_4$ and $V_6$ potentials discussed above, we introduce
the following logarithmic potential 
\begin{align} 
V(\phi)=V_0+B \phi^2+C \phi^4\left(\log\frac{g^2\phi^2}{\overline{v}^2} -\frac{1}{4}\right).
\end{align}
This potential arises in classically scale-invariant models, and its thermal evolution typically exhibits strongly supercooled phase transitions \cite{Jinno:2016knw, Lewicki:2020jiv}. Equivalently (see Appendix \ref{app:Potentials:Log}) we can consider the dimensionless potential
\begin{align} \label{eq:VlogScaled}
V_{\log}(\phi)=(1-\alpha)+\alpha \phi^2- \phi^4\left[(\alpha-2)\log \phi^2+1  \right],
\end{align}
where $\alpha\in\left(0,1\right)$ and the factor in-front of the action is $\beta=\frac{(2-\alpha)(1-\alpha)}{C}$ in $d = 4$.
We also rewrite the trigonometric potential from equation \eqref{eq:CosinePot} in a dimensionless form by using the result in Appendix \ref{app:Potentials:Cos}:
\begin{align} \label{eq:VcosScaled}
V_{\cos}(\phi)=(1-\alpha)\left[\cos(\phi)-1\right]+\frac{1}{2} \sin^2(\phi),
\end{align}
where $(1-\alpha)=1/\lambda^2$ and the prefactor is $\beta=\frac{v^2}{m^2}(1-\alpha)$ in $d=4$.

\begin{figure}[t]
\begin{subfigure}{.5\linewidth}
\centering
\includegraphics[width=1\textwidth]{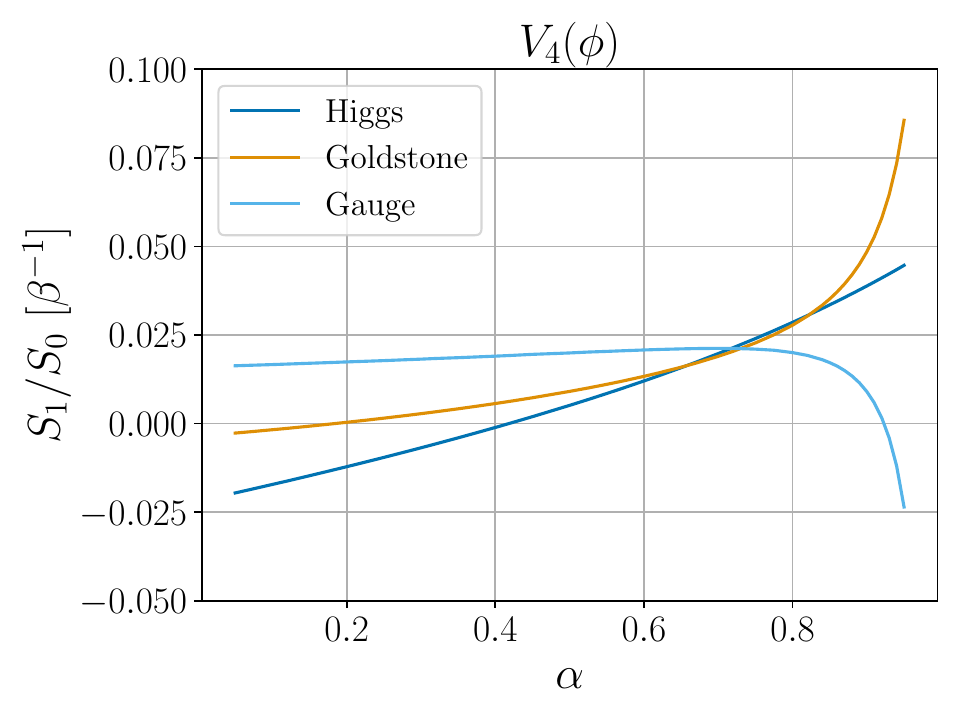}
\caption{}
\label{fig:V_4}
\end{subfigure}%
\begin{subfigure}{.5\linewidth}
\centering
\includegraphics[width=1\textwidth]{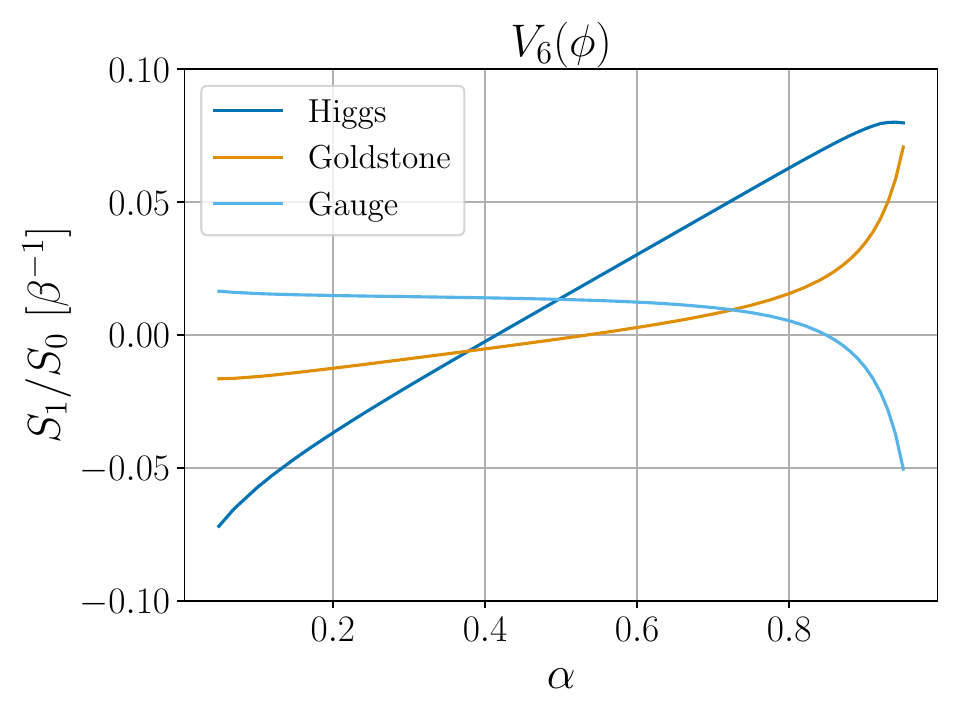}
\caption{}
\label{fig:V_6}
\end{subfigure}\\
\begin{subfigure}{.5\linewidth}
\centering
\includegraphics[width=1\textwidth]{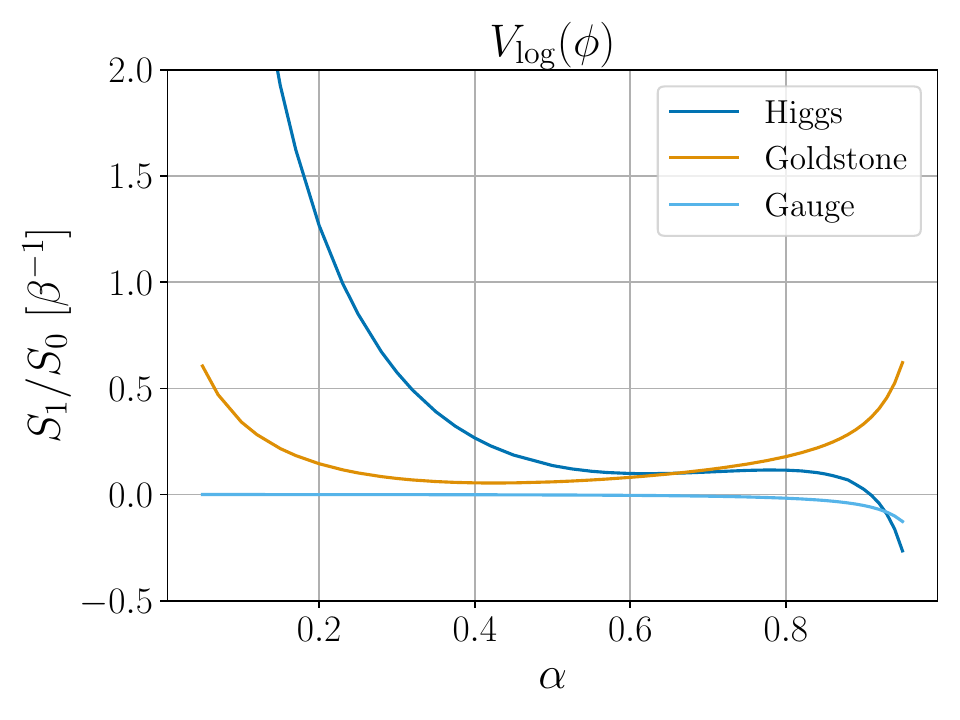}
\caption{}
\label{fig:V_log}
\end{subfigure}%
\begin{subfigure}{.5\linewidth}
\centering
\includegraphics[width=1\textwidth]{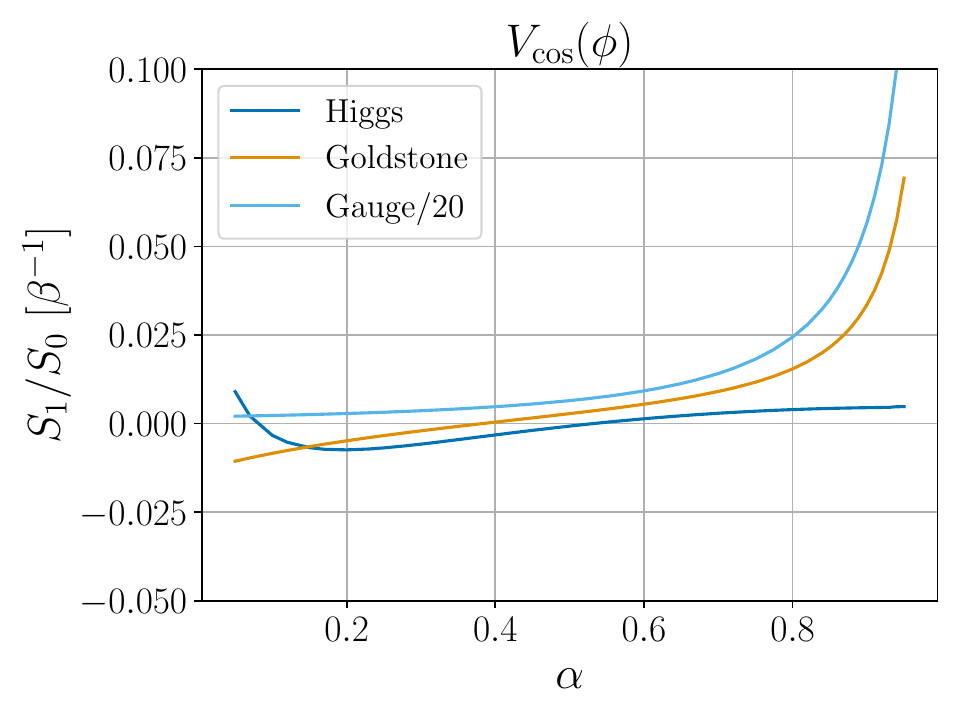}
\caption{}
\label{fig:V_cos}
\end{subfigure}
\caption{Size of Higgs, Goldstone, and vector determinants for various potentials. Figure \ref{fig:V_4} corresponds to the potential in equation \eqref{eq:V4Scaled}; figure \ref{fig:V_6} corresponds to the potential in equation \eqref{eq:V6Scaled}; Figure \ref{fig:V_log} corresponds to the potential in equation \eqref{eq:VlogScaled}; and Figure \ref{fig:V_cos} corresponds to the potential in equation \eqref{eq:VcosScaled}. All figures use the dimensionless gauge-coupling $g=0.5$. In addition, the unit of all plots is in $\beta^{-1}$; so increasing $\beta$ from $1$ to $2$ scales all curves by a factor of $\frac{1}{2}$. Note that the scales are different for each plot.}
\label{fig:V_Examples}
\end{figure}

All the potentials we consider are collected in Appendix \ref{app:Potentials}, together with their scaled dimensionless forms. In Figures \ref{fig:V_4}, \ref{fig:V_6}, \ref{fig:V_log}, and  \ref{fig:V_cos} we show the magnitude of the functional determinants for each potential relative to the corresponding tree-level bounce actions, focusing on $d=4$. Note that we have scaled out $\beta$, as well as any group-theoretic factors from symmetry breaking. We see that the behavior, and size, of the functional determinants varies greatly depending on the potential. This is no surprise, as the bounce solution, and the corresponding functional determinants, explore the global properties of the potential, not just the minima.

Notably, for the $V_4$ and $V_\text{cos}$ potentials, the ratio $S_1/S_0$, for the Higgs determinant, tends towards a constant in the thin-wall limit ($\alpha\rightarrow 1$), while it grows as $(1-\alpha)^{-1}$ for the $V_6$ and $V_\text{log}$ potentials. This is because in the thin-wall limit the leading behavior is determined by the (free-energy) pressure difference between the false and true vacua: $-\frac{1}{2}\log \det\mathcal{O}_H\sim \Delta p R^{d} + \mathcal{O}(R^{d-1})$. The quartic and cosine potentials are special because $V''(\phi)$, and thus the pressure, coincide in the two phases at the critical temperature, due to a symmetry between the two phases. So, similarly to the tree-level action, $\log\det\mathcal{O}_H$ behaves as $R^{d-1}$. For the other potentials, and particles, this is not the case. And in general all determinants scale as $R^{d}$; the coefficient can even be found from the one-loop effective potential.

\section{Conclusions} \label{sec:Conclusions}

\bd\ enables computing functional determinants for a wide range of applications.
This is particularly important for the quantitative reliability of studies of cosmological phase transitions. The common approximation $A\approx T^4$ for the nucleation prefactor (see equation~\eqref{eq:intro_saddlepoint}) is the origin of one of the largest sources of theoretical uncertainty in current predictions of the gravitational wave signal, as revealed through residual renormalisation scale dependence \cite{Croon:2020cgk, Gould:2021oba}. \bd\ marks a step in overcoming this theoretical uncertainty, in preparation for planned gravitational wave observatories such as LISA \cite{LISA:2017pwj}.

The code has been thoroughly tested, and reproduces a large number of results found in the literature. It is robust, and has been shown to yield sub-percent errors from the thick to the thin-wall limits and for a wide range of potentials. There is rarely need to tweak meta-parameters. All the results shown in the present paper were computed using default method and tolerance options.%
\footnote{Note however that relatively accurate bounce profiles are typically required as input.}

\bd\ can be installed just as any other Python library, and is available in the PyPi and conda-forge repositories. It is easy to interface with \ct, so that existing scripts using this library can be straightforwardly upgraded to include functional determinants. Computing a functional determinant typically takes less than about a second on a standard laptop, fast enough to do so for parameter scans of models beyond the Standard Model.

The most important extensions for \bd\ are to allow for multiple background fields, and for couplings between fluctuating field degrees of freedom, i.e.~off-diagonal quadratic terms in the action. These extensions are planned for a second version of \bd, and would allow one to tackle, for example, multi-scalar extensions of the Higgs sector.

\section*{Acknowledgements}
The authors would like to thank Aleksandar Ivanov, Marco Matteini, Miha Neme\v{v}sek and Lorenzo Ubaldi for communication regarding the results of Ref.~\cite{Ivanov:2022osf}. O.G.\ would like to thank Lois Overvoorde for advice on software development. We acknowledge the support of the European Consortium for Astroparticle Theory in the form of an Exchange Travel Grant. The work of A.E.\ has been supported by the Swedish Research Council, project number VR:$2021$-$00363$ and by the Deutsche Forschungsgemeinschaft under Germany’s Excellence Strategy - EXC $2121$ Quantum Universe - $390833306$. The work of O.G.\ has been supported by U.K.\ Science and Technology Facilities Council (STFC) Consolidated grant ST/T000732/1, a Research Leadership Award from the Leverhulme Trust and a Dorothy Hodgkin Fellowship from the Royal Society.

\appendix

\section{Potentials used in the paper}\label{app:Potentials}
To compare the effects of different potential functions, we shift our fields and coordinates such that every tree-level action is of the form
\begin{align}
S_\text{Tree-level}(\alpha,\beta)=\beta \tilde{S}(\alpha),
\end{align}
where $\alpha\in\left(0,1\right)$ and $\alpha \to 0_+$ corresponds to the thick-wall limit; and $\alpha \to 1_-$ to the thin-wall one. We further ensure that the scaled tree-level potentials are linear in $\alpha$

In this form the Higgs determinant is independent of $\beta$, up to a $-\frac{1}{2}\log \beta$ contribution from each zero mode. The tree-level bounce action therefore dominates if $\beta \gg 1$, and perturbation theory works well.

\subsection{A \texorpdfstring{$\phi^4$}{quartic} potential}\label{app:Potentials:Phi4}
Consider the potential
\begin{align}
&V_4(\phi)=\frac{1}{2}m^2\phi^2-\frac{1}{2}\eta \phi^3+\frac{1}{8}\lambda \phi^4.
\end{align}
We always have the freedom to re-scale coordinates and to shift our fields. Meaning that we can eliminate three parameters in favour of two dimensionless ratios. For example, in the literature it is common to use
\begin{align}
\phi \rightarrow \frac{m^2}{\eta} \phi, \quad x\rightarrow \frac{1}{m} x,
\end{align}
which means that the action and potential become
\begin{align}
S[\phi]\rightarrow \beta S[\phi],\quad
V_4(\phi)\rightarrow \frac{1}{2}\phi^2-\frac{1}{2}\phi^3+\frac{1}{8}\alpha \phi^4,
\end{align}
where $\alpha=\lambda m^2\eta^{-2}$ and $\beta=m^{6-d}\eta^{-2}$. 

\subsection{A \texorpdfstring{$\phi^6$}{sextic} potential}\label{app:Potentials:Phi6}
Consider
\begin{align}
&V_6(\phi)=\frac{1}{2}m^2\phi^2-\frac{1}{4}|\lambda|\phi^4+\frac{1}{32}c_6 \phi^6.
\end{align}
which, after scaling $x\to m^{-1} x$, $\phi \to m |\lambda|^{-1/2}\phi $, becomes
\begin{align}
S[\phi]\rightarrow \beta S[\phi],\quad
V_6(\phi)=\frac{1}{2}\phi^2-\frac{1}{4}\phi^4+\frac{1}{32}\alpha \phi^6.
\end{align}
with $\alpha=c_6 m^2|\lambda|^{-2}$ and $\beta=m^{4-d}|\lambda|^{-1}$.

\subsection{A logarithmic potential}\label{app:Potentials:Log}
Consider the potential
\begin{align} 
V(\phi)=V_0+B \phi^2+C \phi^4\left(\log\frac{g^2\phi^2}{v^2} -\frac{1}{4}\right).
\end{align}
We perform the following re-definitions
\begin{align}
&x\rightarrow \frac{\overline{v}}{\sqrt{V_0}} \sqrt{(1-\alpha)}x, \quad \phi \rightarrow \overline{v} \phi, v\quad v\rightarrow  \exp\left(\frac{\alpha+2}{16-8\alpha}\right)g \overline{v}.
\end{align}
After which we find the dimensionless potential
\begin{align}
V_{\log}(\phi)=(1-\alpha)+\alpha \phi^2- \phi^4\left[(\alpha-2)\log \phi^2+1  \right],
\end{align}
where
\begin{align}
&\alpha=\left[W\left(- \frac{e^\frac{1}{4}B g^2}{2 C \overline{v}^2} \right)-1/2\right]^{-1}+2,
\\&\beta=\frac{(2-\alpha)(1-\alpha)}{C}\left(\frac{\alpha}{B} \right)^{(d-4)/2}.
\end{align}
Here $W$ is the Lambert W function and $\beta$ is the overall factor multiplying the action.

\subsection{A trigonometric potential}\label{app:Potentials:Cos}
Take the potential
\begin{align}
V(\phi)=m^2 v^2\left[\cos(\phi/v)-1+\frac{\lambda^2}{2} \sin^2(\phi/v)\right].
\end{align}
To rewrite this potential in our preferred form we use
\begin{align}
x\rightarrow m^{-1} \sqrt{(1-\alpha)} x , \quad \phi \rightarrow v \phi,
\end{align}
to find
\begin{align}
V_{\cos}(\phi)=(1-\alpha)\left[\cos(\phi)-1\right]+\frac{1}{2} \sin^2(\phi),
\end{align}
where $(1-\alpha)=1/\lambda^2$ and the overall factor is $\beta=v^2 m^{2-d}(1-\alpha)^{\frac{d}{2}-1}$.

\section{Vector fields}\label{app:Vector}
As a minimal example, consider the gauged version of the U(1) model of equation \ref{eq:U1GlobalLagrangian}, the Abelian Higgs model,
\begin{align} \label{eq:AbelianHiggsLagrangian}
\mathscr{L}&=\frac{1}{4}F_{\mu \nu}F_{\mu \nu}
+ (D_\mu \Phi)^*  D_\mu\Phi
+ V(\Phi),
\end{align}
where $F_{\mu\nu} = \nabla_\mu A_\nu - \nabla_\nu A_\mu$ and $D_\mu \Phi=\nabla_\mu \Phi-i g A_\mu \Phi$. We choose the class of Fermi gauges,
with gauge-fixing parameter $\xi$ and ghost field $c$\footnote{\newtext{The Nielsen identity~\cite{Nielsen:1975fs} ensures that the full functional determinant is gauge-independent, yet in practice this can be subtle; see Refs.~\cite{Garny:2012cg,Plascencia:2015pga,Lofgren:2021ogg,Hirvonen:2021zej,Endo:2017tsz,Endo:2017gal,Baacke:1999sc} for discussions at both zero and at finite temperature.}}.

If we again assume a radially-symmetric background $\phi(r)$, and expand in fluctuations about this, we find%
\begin{align} \label{eq:AbelianHiggsExpanded}
\mathscr{L}=
&\mathscr{L}(\phi)
+\frac{1}{2}A_\mu \big[(-\nabla^2 + g^2 \phi(r)^2)\delta_{\mu \nu}+\left(1-\xi^{-1}\right)\nabla_\mu \nabla_\nu \big]A_\nu
\nonumber \\
&+\frac{1}{2}H \left[-\nabla^2+V''(\phi)\right]H 
+\frac{1}{2}G\left[-\nabla^2+\phi^{-1}V'(\phi)\right] G
+\bar{c}\left[-\nabla^2\right] c
\\
&+g\nabla_\mu \phi(r)  A_\mu G - g \phi(r) A_\mu  \nabla_\mu G+\ldots \nonumber
\end{align}
For this model, the quadratic part of the action is not diagonal in this field basis, as one can see from the terms on the last line.

The first version of \bd\ is not able to accommodate such off-diagonal terms. However, counting physical degrees of freedom, one expects roughly that the result should be expressible in terms of $d-1$ massive vector degrees of freedom, one Goldstone degree of freedom, and one scalar.%
\footnote{The ghost field drops out in Fermi gauges, as its functional determinant is independent of the background field.}
This counting is apparent in the one-loop Landau-gauge effective potential $V_{1}(\phi)$, for which the off-diagonal derivative terms are zero,
\begin{align}
    e^{-\int_x V_{1}(\phi)} &=
    \det\mathcal{O}_{A_\mu}(\phi)^{-(d-1)/2}
    \det\mathcal{O}_{G}(\phi)^{-1/2}
    \det\mathcal{O}_{H}(\phi)^{-1/2},
\end{align}
where we have defined
\begin{align}
    \mathcal{O}_{A_\mu}(\phi_\text{F}) \equiv -\nabla^2 + g^2 \phi(r)^2.
\end{align}
From this, one expects
\begin{align} \label{eq:RateSymmetryBreaking}
\Gamma \approx 
\left(\frac{\det\mathcal{O}_{A_\mu}(\phi_\text{F})}{\det\mathcal{O}_{A_\mu}(\phi_\text{b})}\right)^{(d-1)/2}
\mathcal{V}_{G} \mathcal{J}_{G}
\sqrt{\frac{\det\mathcal{O}_{G}(\phi_\text{F})}{\det'\mathcal{O}_{G}(\phi_\text{b})}}
\mathcal{J}_{H}
&
\sqrt{\left|\frac{\det\mathcal{O}_{H}(\phi_\text{F})}{\det'\mathcal{O}_{H}(\phi_\text{b})}\right|}
\nonumber \\
&\qquad\qquad \times
    e^{-(S[\phi_\text{b}]-S[\phi_\text{F}])}.
\end{align}
For the one-loop contribution to the rate, the off-diagonal derivative terms are not small, though we expect that neglecting them should nevertheless give a reasonable indication of the order of magnitude.

For models with larger gauge groups, or more scalars, one can likewise neglect any mixing. The final result is then always in a form similar to equation \eqref{eq:RateSymmetryBreaking}.

\section{Algorithm details}
\subsection{Initial value problem}\label{app:ivp}
For the initial value problem in the application of the Gelfand-Yaglom method, one must be mindful to avoid $1/r$ singularities at the origin. We integrate the first step of the initial value problem, from $r=0$ to $\delta r$, using
\begin{align}
    \partial^2 T_l(0) &= \frac{\Delta W (0)}{(d+2l)}, \qquad\qquad
    \partial^3 T_l(0) = 0,\\
    \partial^4 T_l (0) &= \frac{3}{(d+2l+2)} \left[\frac{-4 W(\infty)\Delta W(0)}{(d+2l)^2} + \frac{\Delta W(0)^2}{(d+2l)} + \frac{V'(\phi_\text{b}(0)) \frac{dW}{d\phi}\big|_{r=0}}{d}\right],
\end{align}
so that
\begin{align}
    T_l(\delta r)  &= 1 + \frac{1}{2}\partial^2 T_l(0) \delta r^2 + \frac{1}{4!} \partial^4 T_l(0) \delta r^4, \\
    \partial T_l(\delta r) &= \partial^2 T_l(0) \delta r + \frac{1}{3!} \partial^4 T_l(0) \delta r^3.
\end{align}
After the first step, there are no further coordinate singularities, and we update with the fourth order Runga-Kutta algorithm.

\subsection{Gelfand-Yaglom method for massless bounces} \label{app:massless_algorithm}
Bounces that behave as $\phi_\text{b}\sim r^{-(d-2)}$ for large $r$ require special treatment. In this case the Gelfand-Yaglom solution converges slowly, and to get a decent approximation the grid must be pushed to ever-larger $r$. To circumvent this in \bd\ we instead choose a modest $r_\text{max}$ for which $\phi_\text{b}$ behaves as $\phi_\text{b}(r_\text{max})\sim (r_\text{max})^{-(d-2)}$. We then solve equation \eqref{eq:GYDiffEq} numerically from $r=0$ to $r=r_\text{max}$, and use the solution at $r=r_\text{max}$ to solve equation \eqref{eq:GYDiffEq} analytically from $r=r_\text{max}$ to $r=\infty$. The asymptotic behaviour of $\Delta W$, defined in equation  \eqref{eq:DeltaW}, depends on the form of the potential. To handle general cases we perform a fit to $\Delta W\approx W_\infty r^{-a_\infty}$ for large $r$, for which equation \eqref{eq:GYDiffEq} can be solved analytically. This procedure significantly improves convergence even with a small $r_\text{max}$. We similarly improve the WKB approximation by performing various integrals analytically from $r=r_\text{max}$ to $r=\infty$.

\subsection{Algorithm for fitting \texorpdfstring{$\phi_\infty$}{phi infinity}}\label{app:phiInf}
\begin{figure}[t]
    \centering
    \includegraphics[width=0.618\textwidth]{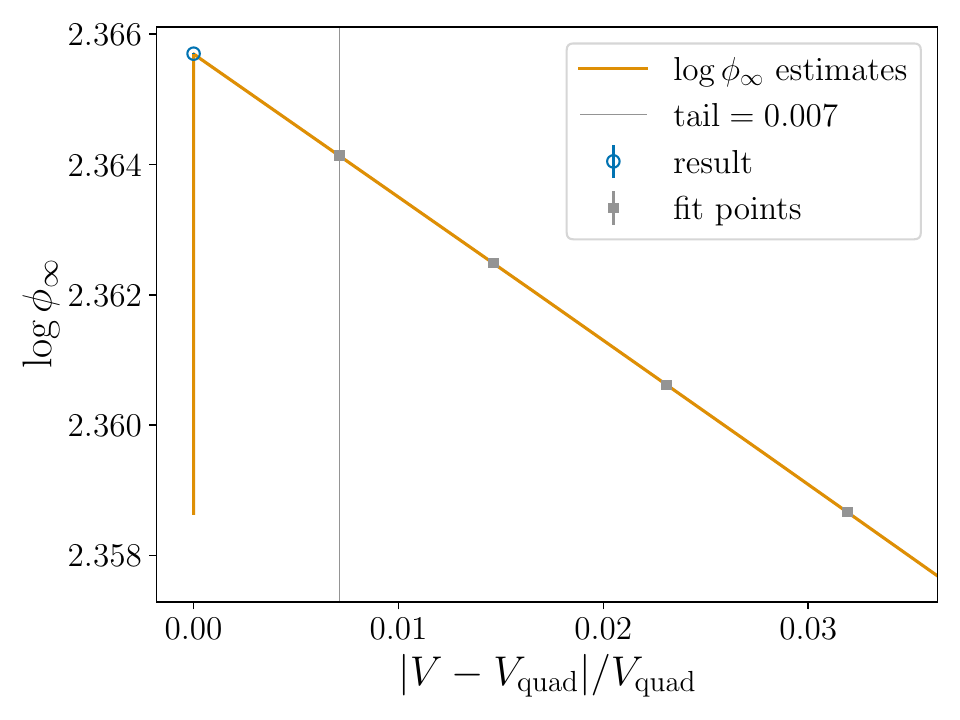}
    \caption{The linear behavior of the \logphiinf\ estimates. The result is obtained from a linear extrapolation from the fit points, which are chosen according to the \tail\ parameter. The failure of estimates happens near zero. Error estimates are plotted but not visible.}
    \label{fig:lnPhiInfLinear}
\end{figure}

The basic idea in the main method for the massive case is to find a quantity for which the direct estimates of \logphiinf\ behave linearly. This way, one can use the linearity to extrapolate to infinite radius. The algorithm uses the following quantity,
\begin{equation}\label{eq:proDevFromQuadPot}
    \frac{\Delta V_\text{quad}}{V_\text{quad}}(\phi)
    \equiv
    \frac{\abs{V(\phi)-V(\phi_\text{F})-\tfrac{\mF^2}{2}\phi^2}}{\tfrac{\mF^2}{2}\phi^2}\,.
\end{equation}
Note that the possible numerical inaccuracies of the bounce solution at large radii are packed very near zero in terms of $\Delta V_\text{quad}/V_\text{quad}$. This follows from the approximately logarithmic relation with the radius,
\begin{equation}
    r\sim-\frac{c}{m_\text{F}}\log\frac{\Delta V_\text{quad}}{V_\text{quad}} \,,
\end{equation}
for some $c=\mathcal{O}(1)$ depending on the form of the potential. See Figure~\ref{fig:lnPhiInfLinear} for an example.

The algorithm chooses four points (shown as gray squares in Figure~\ref{fig:lnPhiInfLinear}), which are then used to extrapolate to
\begin{equation}
    \frac{\Delta V_\text{quad}}{V_\text{quad}}=0\,.
\end{equation}
This corresponds to taking the $r\to\infty$ limit.

The four points are chosen according to the two user-given parameters, with default values \texttt{tail=0.007} and \texttt{log\_phi\_inf\_tol=0.001}. These values were tested to be stable across different models. 

In an idealised case without any numerical inaccuracies, the four points are chosen to have values \tail, 2 \tail, 3 \tail\ and 4 \tail, as illustrated in Figure~\ref{fig:lnPhiInfLinear}. However, to account for numerical inaccuracies in the profile and potential, the algorithm performs consistency tests, which may lead to choosing different points.

Starting from $r=r_\text{max}$ and working inwards towards $r=0$, the algorithm chooses the subset of points that pass the following criteria:
\begin{itemize}
    \item Both $|\phi-\phi_\text{F}|$ and $\Delta V_\text{quad}/V_\text{quad}$ must be larger than the next point away from the centre of the profile.
    \item The floating point errors due to the subtraction in equation~\eqref{eq:proDevFromQuadPot} are very small in comparison to \tail.
\end{itemize}
If any of these conditions fail, the algorithm discards that and all points with larger $r$. Then, it begins again from the next point, working inwards.

Once a point is deemed eligible, the algorithm chooses it as the first chosen point $\phi_1$ if $\Delta V_\text{quad}/V_\text{quad} > \tail$. The rest of the points are chosen if they are above the \tail\ value set by the previous chosen point $\phi_i$:
\begin{equation}\label{eq:pointSpread}
    \frac{\Delta V_\text{quad}}{V_\text{quad}}(\phi_{i+1})>(i+1)\, \text{\tail}_{i}\equiv \frac{i+1}{i}\frac{\Delta V_\text{quad}}{V_\text{quad}}(\phi_{i})\,.
\end{equation}
In addition, the second chosen point $\phi_2$ has to have a small enough estimated error due to the proximity of the end of the profile. One can estimate this as
\begin{equation} \label{eq:PhiInfinityErrorEstimates}
    \abs{\log\Big\{ 1 - \exp\big[-2 \mF (r_{\text{max}}-r)\big] \Big\}}\,.
\end{equation}
This would be exact if the tail satisfied the linearised equation of motion exactly with the boundary condition that $\phi(r_{\text{max}})=\phi_{\text{F}}$. For the second chosen point, we ensure that this is less than \lpit, as a relative error.
Overall, this procedure improves the stability of the extrapolation to numerical inaccuracies in the profile or potential.

The algorithm then extrapolates to infinite radius, performing both a quadratic and a linear fit, and using equation \eqref{eq:PhiInfinityErrorEstimates} to normalise the $\chi^2$. The final error for the main method is chosen to be the maximum of two estimates: the difference between linear and quadratic fits, and the square root of the covariance in the linear fit.

Note that when improving the accuracy of the numerical critical bubble while keeping \tail\ and \lpit\ fixed, the error for the resulting \logphiinf\ does not converge to zero, but rather to a constant. To shrink the errors further requires shrinking the \tail\ and \lpit\ parameters, in addition to improving the bubble. However, these errors in \logphiinf\ are typically overshadowed by other sources of errors in the computation of the full determinant.

The algorithm for the massless case is quite different. The power-like behaviour of the asymptotic field makes a direct fit in $r$ more feasible than the massive case. However, errors from the tail-end of a bubble profile only decrease as a power of $(r_\text{max}-r)$ and not exponentially. The algorithm performs a fit to
\begin{equation}
    \phi(r) \overset{r\to\infty}{\longrightarrow} \phi_\text{F} + \phi_\infty2^{d/2-2}\Gamma(d/2-1) / r^{d-2} + \Delta\phi.
\end{equation}
Here, the constant $\Delta\phi$ is the other solution to the linearised equation of motion around the metastable phase. It does not adhere to the boundary condition and hence picks up deviations from the correct asymptotic behavior.

We estimate the error based on two sources: The square root of the covariance in the fit, $\Delta_\text{cov}\phi_\infty$, and also the error $\Delta_\text{const}\phi_\infty$ from the constant $\Delta\phi\neq 0$ being nonzero.
The total error estimate for \logphiinf\ is then
\begin{equation}
    -\log\Big(1-\frac{\Delta_\text{cov}\phi_\infty+\Delta_\text{const}\phi_\infty}{\phi_\infty}\Big)\,.
\end{equation}
This can become infinite or complex, for example when the tail crosses the initial phase at $r_\text{max}$. In these cases, the fit is discarded, and the result of the fail-safe algorithm, described in Section \ref{sec:Code}, is used instead. More generally, the result with the smaller error is returned.

\subsection{Algorithm for finding the negative eigenvalue}\label{app:NegativeEig}
Here we discuss further details of the implemented discretization.

The first and second derivatives in the differential operator are discretized in $M$ so that their values would be exact if $f_i$ was a fourth-order polynomial around the evaluation point.
This leads to the discretization error for the negative eigenvalue decreasing as $N^{-4}$, when increasing the number of points, $N$.
Hence, the derivative and the second derivative require information from five points of $j$ around the evaluation point, $i$. (Compare with five free parameters in a fourth-order polynomial.) The points are chosen as $j\in\{i-2,\dots,i+2\}$. Thus, the $i$th row of the discretized matrix looks like
\begin{equation}
    M_{ij}=(0,\dots,0,M_{i\,i-2},\dots,M_{i\,i+2},0\dots,0)_i\,.
\end{equation}

The chosen implementation of the derivatives is not straightforwardly possible near the boundaries $i=0$ and $i=N-1$, as the index $j$ would go out of bounds. Near the center of the bubble profile, the matrix $M$ is modified as if the profile would continue beyond $i=0$ as
\begin{align}
    f_{-2}=f_2\,,&\quad f_{-1}=f_{1}\,,
\end{align}
which implements the boundary condition $\partial f = 0$ at the origin. At the end of the profile, the modification of the matrix depends on the chosen boundary condition. It corresponds to the continuation of the profile beyond $f_{N-1}$ as either
\begin{align}
    f_{N}= f_{N-1}\,,\quad\text{or}\quad f_{N}=0\,,
\end{align}
where the former corresponds to the zero derivative and the latter to the zero value.

Finally, we want to discuss one subtle point in the discretisation at $r=0$: The second term in the differential operator in equation \eqref{eq:diffEigValProblem} appears singular. This is however not a problematic singularity. Due to the condition that $\partial f(r)\to0$ as $r\to0^+$, one can show that
\begin{equation}
    \lim_{r\to0^+}\frac{1}{r}\partial f(r)=\partial^2 f(0)\,.
\end{equation}
Thus, for the first row of the matrix $M$, corresponding to $r=0$, one can use the discretized version of $\partial^2$ instead of that of the apparently singular operator
\begin{equation}
     \partial^2+\frac{d-1}{r}\partial \,.
\end{equation}

\section{Volume, Jacobians, and removing zero modes}\label{app:Volume}
\subsection{Breaking translational invariance}
Zero modes arise if the bounce breaks a symmetry of the full theory. In particular, consider small deviations about the bounce:
\begin{align} \label{eq:linear_transform}
\phi(x)=\phi_\text{b}(r)+c^a \xi_a(x),
\end{align}
where $\xi_a(x)$ are a complete set of eigenvectors that we normalize as $\int \mathrm{d}^d x \xi^a \xi^b=2 \pi \delta^{ab}$, and the integration measure is $\Pi_a \mathrm{d} c^a.$

Let us start with translational symmetries. Since the bounce isn't invariant under coordinate shifts $x_\mu \rightarrow x_\mu +a_\mu$, we expect $d$ zero modes; which we denote by $\xi_\mu(x)$. The idea with collective coordinates is that we can equivalently express these zero modes as a coordinate shift of the bounce%
\footnote{Above the one-loop order we have to enlarge this coordinate shift so that it encompasses all eigenfunctions. This gives rise to a more complicated Jacobian.}:
\begin{align}
\phi_\text{b}(x+a)=\phi_\text{b}(r)+a_\mu \nabla_\mu \phi_\text{b}(r)+\ldots
\end{align}
As such we can identify $\xi_\mu(x)=\mathcal{N}^{-1} \nabla_\mu \phi_\text{b}(r)$ and $c_\mu=a_\mu \mathcal{N}$ from equation \eqref{eq:linear_transform}. To fix the normalisation $\mathcal{N}$ we use
\begin{align}
\mathcal{N}^{-2} \int \mathrm{d}^d x \nabla_\mu\phi_\text{b} \nabla_\nu\phi_\text{b}=2\pi \delta_{\mu \nu},
\end{align}
which implies that $\mathcal{N}^2 = S[\phi_\text{b}]/(2\pi)$. The Jacobian for the transformation of the integration measure for all $d$ modes is then $\mathcal{J} = \mathcal{N}^d = (S[\phi_\text{b}]/(2\pi))^{d/2}$

In effect all zero modes are absorbed by the bounce, and the integral over the zero modes yields
\begin{align}
    \int \mathrm{d}^d c=\mathcal{J}\int \mathrm{d}^d a = \mathcal{J} \mathcal{V},
\end{align}
where $\mathcal{V}$ is the spacetime volume.

Translation zero modes arise in the Higgs determinant at $l=1$ in the sum over angular quantum number\te as $\nabla_\mu \phi_b(r)$ transforms as a vector. The full result for integration over modes at $l=1$ is given in equation \eqref{eq:packTransZeroModes}.

\subsection{Translations for massless Higgs potentials} \label{app:massless_zero_modes}
The derivation of equation \eqref{eq:packTransZeroModes}, for translational zero modes assumes that the nucleating scalar is massive in the false vacuum, $\mF^2=V''(\phi_\text{F})>0$. As discussed around equation \eqref{eq:l1DeformedDifferentialEquation}, this was done
by deforming the $l=1$ differential equation into:
\begin{align}
	\left[-\partial^2-\frac{d-1}{r}\partial +\frac{(d-1)}{r^2}+V''(\phi_\text{b})+k^2\right]\psi^{1,k}_{b}(r)=0.
\end{align}
This shifts all eigenvalues by $k^2$, and as such we can remove the final zero mode by using
\begin{align}
\frac{\det'\left(-\nabla_1^2+V''(\phi_\text{b})\right)}{\det\left(-\nabla_1^2+V''(\phi_\text{F})\right)}=
\lim_{k^2\rightarrow 0}  \frac{\psi^{1,k}_\text{b}(\infty)}{k^2}\frac{1}{\psi^1_\text{F}(\infty)}.
\end{align}
However, this procedure does not have a smooth $\mF\to 0$ limit. To see this, let us follow Ref.~\cite{Dunne:2005rt} and express $\psi^{1,k}_\text{b}(R)$ via
\begin{align}
R^d \left[\partial \psi_\text{b}^1 \psi^{1,k}_\text{b}-\partial \psi^{1,k}_\text{b} \psi_\text{b}^1 \right]_{r=R}=-k^2 \int_0^R \mathrm{d}r r^d \psi_\text{b}^1\psi^{1,k}_\text{b},
\end{align}
where $\psi_\text{b}^1(r)=\frac{\partial \phi_\text{b}(r)}{\partial^2 \phi_\text{b}(0)}$ is the normalized zero mode.
In the massive case, we can then solve for $\psi^{1,k}_\text{b}(R)$ by using that $\psi^{1,k}_\text{b}(r)\sim e^{r\sqrt{\mF^2+k^2}}$ and $\psi_\text{b}^1(r)\sim e^{-r \mF}$ at large $r$. This needs modification for $\mF=0$, as the behaviour of the solutions changes. 

To proceed it is useful to add $k^2$ also for the false-vacuum solution and to directly work with $T_{1,k} (r)=\frac{\psi^{1,k}_\text{b}(r)}{\psi^{1,k}_\text{F}(r)}$. The equation for $T_{1,k}$ is
\begin{align}\label{eq:TZeroMode}
&\left[-\partial^2 -U_{1,k} \partial +V''(\phi_\text{b})\right] T_{1,k} =0,\quad \partial T_{1,k}(0)=0,\quad T_{1,k}(0)=1,
\\&U_{1,k} =\frac{d+1}{r}+\frac{2k^2r}{d+2}+\mathcal{O}(k^4r^3),
\end{align}
where $U_{1,k}$ is the consequent $k^2$ deformation of equation \eqref{eq:UFunction}.
The solution at $k^2=0$ is
\begin{align}
    T_1 \equiv \frac{\partial \phi_\text{b}(r)}{r \partial^2 \phi_\text{b}(0)},
\end{align}
leading to $T_1(\infty) = 0$ as expected.

Now, similarly to Ref.~\cite{Dunne:2005rt} our strategy is to integrate equation \eqref{eq:TZeroMode} with a suitable factor. To get something useful we choose
\begin{align}
\int_0^R \mathrm{d}r r^{d+1} T_{1} \left[\partial^2 T_{1,k} + U_{1,k} \partial T_{1,k}-V''(\phi_\text{b}) T_{1,k} \right]=0.
\end{align}
The $r^{d+1}$ factor is crucial for the equation to solve: It is designed to cancel the $\frac{d+1}{r}$ term in $U_{1,k}$, when integrating by parts. We can now move derivatives from $T_{1,k}$ to $T_{1}$ to arrive at
\begin{align}
    R^{d+1}\left[\partial T_1T_{1,k} - T_1 \partial T_{1,k}\right]_{r=R} + \int_0^R \mathrm{d}r r^{d+1} T_{1} \frac{2k^2r}{d+2} \partial T_{1,k} = 0.
\end{align}
Taking $R$ much larger than any intrinsic scale from $\phi_\text{b}$, so that the $V''(\phi_\text{b})$ term can be dropped from equation \eqref{eq:TZeroMode}, we find the large $r$ asymptotics,
\begin{align}
T_1(r)&\sim \frac{-2^{d/2-1}\Gamma\left(\frac{d}{2}\right)\phi_\infty}{\partial^2 \phi_\text{b}(0)} r^{-d},
& T_{1,k}(r) &\sim a + \frac{b}{r^{d+2}}\exp\left({-\frac{k^2r^2}{d+2}}\right)+T_1(r),
\end{align}
for some constants $a$, $b$. Using this we can solve for $T_{1,k}(\infty)$ to leading order in $k^2$:

\begin{align}
T_{1,k}(\infty)&=\frac{-k^2 \partial^2 \phi_\text{b}(0)}{2^{d/2}\Gamma\left(\frac{d}{2}+1\right)  \phi_\infty }\int_0^\infty \mathrm{d}r r^{d+1}(T_{1})^2,
\\&=\frac{-k^2}{2^{d/2}\Gamma\left(\frac{d}{2}+1\right) \phi_\infty \partial^2\phi_\text{b}(0)}\int_0^\infty \mathrm{d}r r^{d-1} \partial\phi_\text{b}(r)^2, 
\\&=\frac{-k^2}{(2\pi)^{d/2} \phi_\infty \partial^2\phi_\text{b}(0)} S[\phi_\text{b}].
\end{align}
All in all we find
\begin{align}
\left(\frac{S[\phi_\text{b}]}{2\pi}\right)^{d/2}	\left(\frac{\det'\left(-\nabla_1^2+V''(\phi_\text{b})\right)}{\det\left(-\nabla_1^2+V''(\phi_\text{F})\right)}\right)^{-d/2}=\left[(2\pi)^{d/2-1}\phi_\infty \abs{\partial^2\phi_\text{b}(0)}\right]^{d/2},
\end{align}
the same expression as in the massive case.
As an example, take $d=4$ and $V(\phi)=-\frac{1}{4} \lambda\phi^4$; the Fubini-Lipatov instanton is then~\cite{Fubini:1976jm,Lipatov:1976ny}
\begin{align}
\phi_\text{b}(r)=\sqrt{\frac{8}{\lambda}}\frac{R}{R^2+r^2},
\end{align}
where $R>0$ is an arbitrary parameter. With this ``bounce'' we identify
\begin{align}
\phi_\infty=\sqrt{\frac{8}{\lambda}} R, \quad \abs{\partial^2\phi_\text{b}(0)}=\sqrt{\frac{32}{\lambda}} R^{-3},
\end{align}
and thus
\begin{align}
\left(\frac{S[\phi_\text{b}]}{2\pi}\right)^{2}	\left(\frac{\det'\left(-\nabla_1^2+V''(\phi_\text{b})\right)}{\det\left(-\nabla_1^2+V''(\phi_\text{F})\right)}\right)^{-2}=\left(\frac{32 \pi}{ \lambda R^2}\right)^2,
\end{align}
which is in agreement with~\cite{Andreassen:2017rzq, Isidori:2001bm}.

\subsection{Internal symmetries} \label{app:GoldstoneZeroModes}
We next consider Goldstone bosons. In this case our theory is invariant under the action of some internal symmetry group $\mathcal{G}$; we write the linearised action of this group on our scalars as
\begin{align}
\phi_i\rightarrow \phi_i + \varepsilon_a T^a_{ij} \phi^j,
\end{align}
where $a$ runs over the group generators, $1, \dots, \text{dim}(\mathcal{G})$, and $i, j$ run over the indices of the representation in which $\phi$ transforms.
The bounce will generically only be invariant under a subset of these generators $T\in \mathcal{H} \subset \mathcal{G}$, while the other, broken ones, satisfy
\begin{align}
\epsilon_{\overline{a}}T^{\overline{a}}_{ij} \phi^j_\text{b}(r) \neq 0,
\end{align}
where we have introduced barred indices $\overline{a}$ to run over the broken generators $T^{\overline{a}}\in \mathcal{G}/\mathcal{H}$. In general then we expect $\text{dim}(\mathcal{G}/\mathcal{H})$ zero modes. To remove these zero-modes we again express the linear shift in terms of a basis of functions
\begin{align}
\phi^i(x)=\phi^i_\text{b}(r)+c^{\overline{a}} \xi_{\overline{a}}^{i}(x),
\end{align}
The broken generators rotate one solution, $\phi^i_\text{b}$, to another solution.
So we can identify $\xi^{i}_{\overline{a}}(x)=\mathcal{N}_\text{G}^{-1}T^{\overline{a}}_{ij}\phi^j_\text{b}(r)$ and $c^{\overline{a}}=\mathcal{N}_\text{G} \varepsilon^{\overline{a}}$.
The normalization factor for the Goldstones is 
\begin{align}
   \mathcal{N}_\text{G}^2 = \frac{1}{2\pi}\int \mathrm{d}^d x \phi_\text{b}^i \phi_\text{b}^j \delta_{ij},
\end{align}
and the corresponding Jacobian is $\mathcal{J}_G = \mathcal{N}_\text{G}^{n_G}$ where $n_\text{G}={\text{dim}(\mathcal{G}/\mathcal{H})}$ is the number of Goldstone bosons.
By absorbing these zero modes as an arbitrary rotation of $\phi_\text{b}^i$ we are left with an integration over the broken subgroup:
\begin{align}
\int \Pi_{\overline{a}} \mathrm{d} c^{\overline{a}} &=
\mathcal{J}_\text{G}
\int \mathrm{d} \Pi_{\overline{a}} \epsilon^{\overline{a}}
= \mathcal{J}_\text{G} \mathcal{V}_\text{G},
\end{align}
where $\mathcal{V}_\text{G}=\text{vol}(\mathcal{G}/\mathcal{H})$ is the volume of the broken subgroup. So for example, a $\mathcal{U}(1)$ group that is broken completely gives $\mathcal{V}_G=2 \pi$. Note that $\mathcal{V}_\text{G}$ depends on the group manifold, and not just the algebra, so it is sensitive to the global structure of the gauge group, such as division by a discrete group; see for example Refs.~\cite{Buchmuller:1993bq, Tong:2017oea}.

These Goldstone zero modes arise at $l=0$ in the sum over angular quantum number. To compute the full $l=0$ result, we can follow the method of Ref.~\cite{Dunne:2005rt}, just as for translation zero modes; see equation \eqref{eq:packTransZeroModes}. The result is
\begin{align}
\mathcal{V}_\text{G} \mathcal{J}_\text{G} \left(\frac{\det'\left(-\nabla_0^2+V'(\phi_\text{b})/\phi_\text{b}\right)}{\det\left(-\nabla_0^2+V'(\phi_\text{F})/\phi_\text{F}\right)}\right)^{-n_\text{G}/2}
=
\mathcal{V}_\text{G} \left[(2\pi)^{d/2-1}\phi_\infty \phi_\text{b}(0)\right]^{n_\text{G}/2}.
\end{align}

\subsection{Dilatations}\label{app:ConformalVolume}
In models with classical scale invariance there is a zero mode arising at $l=0$ from the breaking of scale transformations or dilatations, $\phi(x)\rightarrow s^{\frac{d-2}{2}}\phi(x s)$, $s>0$. For an infinitesimal transformation we take $s=1+\varepsilon$ with $|\varepsilon|\ll 1$. In this case the zero mode is $\xi_s(r)=\frac{d-2}{2}\phi_\text{b}(r)+r \partial \phi_\text{b}(r)$, and the Jacobian factor for the transformation to the collective coordinate is
\begin{align}
\mathcal{J}_s=\left(\frac{\int d^d x \xi_s^2}{2\pi}\right)^{1/2}.
\end{align}
Formally $\mathcal{J}_s$ is infinite\te the zero mode is not normalisable\te but the full determinant will be finite once we remove the zero mode \cite{Andreassen:2017rzq}.

Like the integration over translations, the integration over scale factors $\mathcal{V}_s = \int_0^\infty \mathrm{d}s$ is divergent in a standard saddle-point evaluation of the path integral. As a consequence, we omit the factor of $\mathcal{V}_s$ in \bd. However, unlike translation invariance, scale invariance often does not survive quantisation, being broken by loop corrections. One may therefore obtain a finite result by integrating over the scale factor after having computed the functional determinant for the other modes \cite{Andreassen:2017rzq}. As the instanton itself breaks scale invariance, practically this requires computing the functional determinant for each value of the scale factor and integrating the result.

Proceeding as before, we deduce,
\begin{align}\label{eq:ScalelessZM}
\mathcal{J}_s
    \left\vert
        \frac{\det'\left(-\nabla_0^2 + V''(\phi_\text{b})\right)}{\det\left(-\nabla_0^2+V''(\phi_\text{F})\right)}
    \right\vert^{-1/2}
=
\left[
    \frac{d-2}{2} (2 \pi)^{d/2-1} \phi_\infty \xi_s(0)
\right]^{1/2}.
\end{align}

The result in equation \eqref{eq:ScalelessZM} is finite, and for the $V(\phi)=-\frac{1}{4} \lambda\phi^4$ potential in $d=4$ we find
\begin{align}
\mathcal{J}_s\left\vert\frac{\det'\left(-\nabla_0^2+V''(\phi_\text{b})\right)}{\det\left(-\nabla_0^2+V''(\phi_\text{F})\right)}\right\vert^{-1/2}
=
\left(\frac{16 \pi}{\lambda}\right)^{1/2},
\end{align}
in agreement with~\cite{Andreassen:2017rzq}.

Finally, we note that in the literature the dilatation collective coordinate is sometimes chosen differently. For the $V(\phi)=-\frac{1}{4} \lambda\phi^4$ potential in $d=4$, it is often chosen as the scale $R$ in the Fubini-Lipatov instanton,
\begin{align}
\phi_\text{b}(r)=\sqrt{\frac{8}{\lambda}}\frac{R}{R^2+r^2}.
\end{align}
With $R$ as the collective coordinate, the dilatation integration measure is $\int \mathrm{d}R$, while with our choice of collective coordinate, $s$, it is $\int \mathrm{d} s = \int \mathrm{d}R / R$. The total integrated results must agree, so the integrand with $s$ as the collective coordinate is $R$ times the integrand with $R$ as the collective coordinate.

\subsection{Special conformal transformations}
In addition to dilatations, we can also have zero modes associated with special conformal transformations:
\begin{align}
\phi \rightarrow \phi+(d-2)b\cdot x \phi-x^2 b\cdot \partial\phi+2 x\cdot b x\cdot \partial \phi.
\end{align}
These zero modes appear in the $l=1$ determinant\te as they are vectors\te and after projecting we find
\begin{align}
\psi^{sc}=(d-2)r\phi_\text{b}(r)+r^2 \partial\phi_\text{b}=r^{4-d} \partial\left[r^{d-2}\phi_\text{b}(r) \right].
\end{align}\label{eq:SpecialConformal}
For $d=4$ we can see from the analytic ``bounce'' that $\psi^{sc}\propto \partial \phi_\text{b}$; so no new zero modes appear. This also holds for conformal bounces in $d=3$ and $d=6$.

\section{Higher-order WKB approximations} \label{app:WKB}
When finding the determinant with the Gelfand-Yaglom method, we have to solve the following equation
\begin{align}
\partial^2 \psi+\frac{d-1}{r}\partial \psi-\frac{l(l+d-2)}{r^2}\psi-W(r)\psi=0,
\end{align}
for $\psi$, with the boundary condition $\psi\sim r^{l}$ for small $r$. We use R.E.~Langer's approach~\cite{Langer:1937qr} and define $\psi=r^{1-d/2}\Psi(x)$ together with a change of variables to $r=e^x$. Equation \eqref{eq:GYDiffEq} is then equivalent to
\begin{align}
	\partial_x^2\Psi^l_{b,F}(x)=A^2(x)\Psi^l_{b,F}, \quad A^2(x)=e^{2x}W(e^x)+\lb^2,\quad \lb \equiv l + \frac{d-2}{2}.
\end{align}
We can now solve this equation with a WKB approximation. To do this, we use that 
$\frac{d}{dx}A(x)\sim \lb^{-1}$. Then, denoting our expansion parameter by $\varepsilon \sim \lb^{-1}\ll 1$, we can write our equation and solution as
\begin{align}
	\varepsilon^2\Psi''(x)=A^2(x)\Psi(x), \quad \Psi(x)=\exp\left[\varepsilon^{-1}S_0+S_1+\varepsilon S_2+\varepsilon^2 S_3 +\ldots\right]
\end{align}
in terms of the undetermined functions $S_n(x)$. The first two orders give
\begin{align}
(S_0)'=\pm A(x), \quad S_1=-\frac{1}{2}\log |A|.
\end{align}

When calculating the integral we have to evaluate
\begin{align}
\log \frac{\Psi_\text{b}(\infty)}{\Psi_\text{F}(\infty)}=\varepsilon^{-1} \Delta S_0(\infty)+\Delta S_1(\infty)+\ldots
\end{align}
It turns out that every WKB correction of the form $S_{2n+1}$ vanishes once we subtract the false-vacuum solution. So we only need the even-numbered corrections. To reach $l^{-9}$ we need all terms up to $S_6$. We find

\begin{align}
&S_0=\pm \int dx A(x),
\\&  S_2=\int dx \frac{2 A(x) A''(x)-3 A'(x)^2}{8 A(x)^3},
\\& \!\begin{aligned}[t] 
	S_4=-\int dx\frac{1}{128 A(x)^7}&\left[-8 A(x)^3 A^{(4)}(x)+52 A(x)^2 A''(x)^2+297 A'(x)^4\right.
	\\&\left.+80 A(x)^2 A^{(3)}(x) A'(x)-396 A(x) A'(x)^2 A''(x)\right]\nonumber
\end{aligned}\\
\\& \!\begin{aligned}[t] 
	S_6=\int dx \frac{1}{1024 A(x)^{11}}&\left\lbrace-50139 A'(x)^6-22704 A(x)^2 A^{(3)}(x) A'(x)^3\right.
	\\&\left.+100278 A(x) A'(x)^4 A''(x)\right.
	\\&\left.+12 A(x)^2 A'(x)^2 \left(290 A(x) A^{(4)}(x)-3679 A''(x)^2\right)\right.
	\\&\left.+8 A(x)^3 \left(301 A''(x)^3+A(x) \left(2 A(x) A^{(6)}(x)-49 A^{(3)}(x)^2\right)\right)\right.\nonumber
	\\& \left. -640 A(x)^3 A(x) A^{(4)}(x) A''(x) \right.\nonumber
	\\&\left.+16 A(x)^3 A'(x) \left(694 A^{(3)}(x) A''(x)-21 A(x) A^{(5)}(x)\right)\right\rbrace \nonumber
\end{aligned}\\
\end{align}
Each of these terms can now be expanded in powers of $l^{-1}$, the result is given in the code.

\bibliographystyle{unsrt}
\bibliography{refs}

\end{document}